\documentclass[journal]{IEEEtran} 
\IEEEoverridecommandlockouts

\usepackage{caption}
\usepackage{subcaption}
\usepackage{cite}
\usepackage{amsmath,amssymb,amsfonts}
\usepackage{amsthm}
\usepackage{graphicx}
\usepackage{textcomp}
\usepackage{xcolor}
\usepackage{lipsum}
\usepackage{arydshln}
\usepackage{mwe}
\def\BibTeX{{\rm B\kern-.05em{\sc i\kern-.025em b}\kern-.08em
    T\kern-.1667em\lower.7ex\hbox{E}\kern-.125emX}}
    
\usepackage{lipsum}
\usepackage{mathtools}
\usepackage{cuted}
\usepackage{dsfont}
\usepackage[linesnumbered,ruled]{algorithm2e} 
    
\newtheorem{theorem}{Theorem}

\newtheorem{assumption}{Assumption}

\newtheorem{definition}{Definition}
\newtheorem{lemma}[theorem]{Lemma}

\theoremstyle{remark}
\newtheorem*{remark}{Remark}
\DeclareMathOperator{\tr}{\mathrm{trace}}

\newcommand{\trace}{{\mbox{trace}}}
\newcommand{\diag}{{\mbox{diag}}}
\newcommand{\cov}{\mbox{Cov}}
\begin{document}

\title{\Huge Edge Selection in Bilinear Dynamical Networks\\
\thanks{This work was partially supported by AFOSR Grant FA9550-21-1-0289 (EDS) and ONR Grant N00014-21-1-2431 (ACBO, MS, EDS)}
}

\author{Arthur Castello B. de Oliveira$^{1}$, Milad Siami$^{1}$, and Eduardo D. Sontag $^{1,2}$
\thanks{$^{1}$Department of Electrical \& Computer Engineering,
Northeastern University, Boston, MA 02115 USA
	(e-mails: {\tt\small \{castello.a, m.siami, e.sontag\}@northeastern.edu}).}
\thanks{$^{2}$Departments of Bioengineering,
Northeastern University, Boston, MA 02115 USA.}%
}

\maketitle

\allowdisplaybreaks
\begin{abstract}
We develop some basic principles for the design and robustness analysis of a continuous-time bilinear network where an attacker can manipulate the strength of the interconnections among some of its neighboring nodes. Bilinear networks characterize scenarios where an external input or an attack not only can affect the state of a node but can also affect the strength of the couplings between them. We begin by carefully defining a bilinear network and reviewing relevant tools from the literature on bilinear dynamical systems. In particular, the $\mathcal{H}_2$-norm of bilinear systems is known to have properties analogous to its linear counterpart and provides valuable insights for identifying which edges are most sensitive to attacks. We then formulate the edge protection optimization problem of picking a limited number of attack-free edges. The exact optimization problem is combinatorial in the number of edges, and brute-force approaches show poor scalability. However, we show that the $\mathcal{H}_2$-norm as cost function is supermodular and, therefore, allows for efficient greedy approximations of the optimal solution. In addition to studying this greedy solution, we relax the integer optimization to obtain a lower bound on the optimal solution and propose two other methods that approximate the optimal solution. We illustrate and compare the effectiveness of our theoretical findings via several numerical simulations using benchmark examples.

\end{abstract}


\section{Introduction}

The robust design of control systems against adversarial attacks is crucial for sustainability, from engineering infrastructures to living cells. %
%
%
{
One way to think of the problem of vulnerabilities is to consider a set of interacting dynamic agents, whose behaviors are influenced by flows between them.  These flows might be purely informational, or they could involve physical objects, such as currents in electrical grids, goods being shipped, people migrations (if the ``agents" are cities or countries), infection vectors being unwittingly exchanged, and so forth. Mathematically, there are many ways to formulate rich and interesting problems. Here, we will assume that there is some quantity associated with each agent, whose time evolution depends not only on its own previous values, but also on the past ``flow of information'' from other nearby agents. Such interactions are modeled by a graph, as in Fig. \ref{fig:introduction}, where the agents are represented by the nodes, and the flow of information by the edges. When designing such systems, one might want to harden the design so as to attenuate the effects of interference by an adversary or random failures that might occur and steer us from a desired goal state. In this context. it is natural to ask which designs optimally reduce the vulnerability of the network, that is, how to make our system safer.

\begin{figure}
    \centering
    \includegraphics[trim=50 80 50 60, clip, scale=0.32]{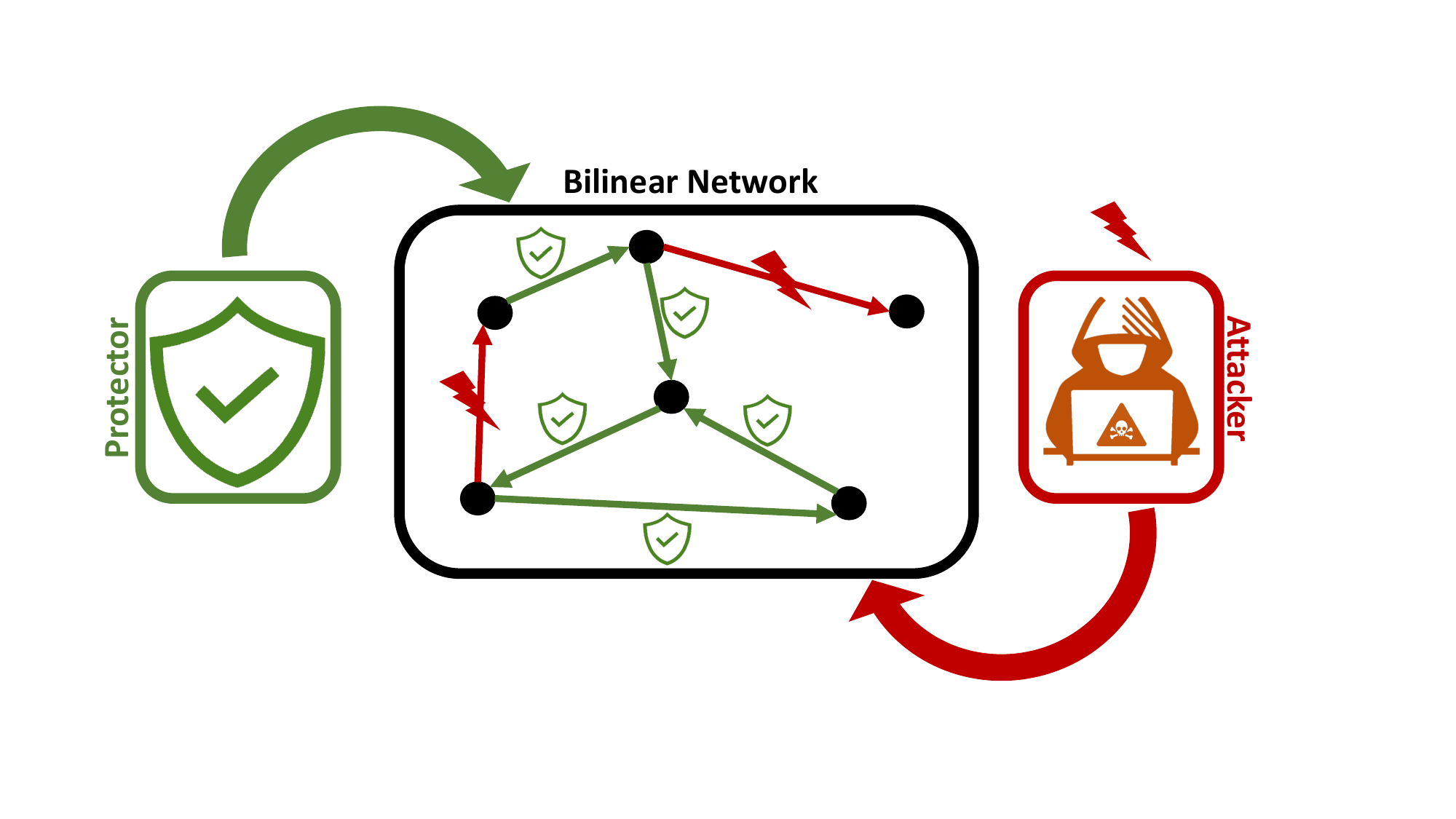}
    \caption{Graphical Representation of a bilinear network with $6$ nodes and $7$ edges. Nodes represent ``agents'' while edges represent ``flow'' between agents. The considered network has an edge protection budget of $5$. The designer decided to protect the $5$ edges painted in green, leaving the $2$ edges in red vulnerable to attack.}
    \label{fig:introduction}
\end{figure}

In the traditional linear dynamic network literature, the actions available to an adversary are restricted to directly affecting the dynamics of specific agents, in the hopes (of the adversary) that these disruptions will propagate to the rest of the network. Even with such a limited assumption, this formulation is still very rich and versatile. Many works can be found in the literature, \cite{shoukry2018smt,chong2015observability,fawzi2014secure,tang2018sensor}, where the authors focus on investigating how disturbances on specific agents propagate and how to evaluate the weak spots of a network.}

Specifically, the performance and robustness analysis of linear consensus networks subject to external stochastic disturbances has been studied in the literature, as in \cite{Bamieh12, siami2016fundamental,Jadbabaie13}, where the $\mathcal H_2$-norm (a well-known measure of robustness for linear systems) was employed as a scalar performance metric.
Supermodularity of a number of control objectives for linear time-invariant (LTI) systems was studied in \cite{summers2015submodularity,summers2015topology}. Specifically, one of the control objectives is the trace of the inverse of the controllability Gramian, which one may interpret as the average control energy required to steer the system to a unit state. In \cite{olshevsky2017non}, it was proved that the average control energy is not always supermodular for LTI systems, contrary to claims in \cite{summers2015submodularity,summers2015topology,8801879}. The importance of such a result lies in the fact that supermodularity of a function would assure us the existence of an efficient approximation algorithm for finding its minimum; that is, if the cost was supermodular, one could solve a minimization problem for large systems with many agents and get a ``good enough'' approximation of the optimum.
The work in \cite{siami2017growing} demonstrates a subclass of differentiable systemic performance measures that are supermodular.

{
Despite this subject having a large literature, a weakness of assuming interference only on the node dynamics is that it restricts our analysis to local conclusions if our adversaries have more power over our network. On the other hand, performing analysis while allowing arbitrary powers for our adversaries is a very hard problem, so in this paper we focus in a compromise. We will assume that our adversaries can also disturb the capacity of the interconnections between agents (in an ``affine'' manner, to be explained), in addition to being able to affect nodes directly.
{
Given a ``protection budget'' $\ell$, we will ask which $\ell$ edges should be protected in order to minimize the overall degradation of performance under an attack.  The first contribution of this paper is to make the problem precise.  We then proceed to study the problem algorithmically.}

In mathematical terms, {
the assumption that edges can be modified in an affine manner translates into} our network having bilinear terms in its dynamics. Even though the tools and intuitions from the linear systems literature do not immediately generalize to bilinear systems, they constitute nonetheless an interesting class of nonlinear systems \cite{Mohler, elliot2009}, since they have universal approximation properties and have been used to model problems in a wide variety of areas ranging from electrical networks to surface vehicles and immunology. } 

Gramian-based reachability metrics for discrete-time bilinear systems were considered in \cite{zhao2016gramian}, where it was shown that the minimum input energy required to steer the state from the origin to any reachable target state can be bounded below by a  Gramian-based reachability metric. Alternatively, to solve model order reduction problems for bilinear systems, in \cite{redmann2019bilinear} the authors show that the $\mathcal{H}_2$ error between the original system and a lower order approximation bounds their output error. These findings explain the previous numerical results where $\mathcal{H}_2$-based model order reduction algorithms provided good approximations. Also related to the subject of this paper, in \cite{benner2017balanced,zhao2015reachability} both the reachability and observability Gramians of bilinear systems are related to lower and upper bounds for the controllability and observability energy functionals, respectively. These results further indicate the usefulness of the $\mathcal{H}_2$-norm as a performance metric for bilinear systems, since it is directly related to the reachability Gramian. 

We formulate the problem of attacking a network through multiplicative disturbances as an optimization problem for a bilinear system. Specifically, the attacker tries to maximize the $\mathcal{H}_2$-norm of the system, while the system designer tries to find the best set of edges to protect. To help solving this issue, we show that, when it is well defined, the $\mathcal{H}_2$-norm of a bilinear system is supermodular on the power set of the vulnerable edges. This property formalizes the intuitive notion that attacks act in synergy and that they are more effective when applied together rather than individually.

We organize the paper as follows: in Section \ref{sec:prelim}, we establish the notation and definitions that lead to our main result, and we also formally define the structure considered for the bilinear systems and networks; in Section \ref{sec:theory}, we present the definition of $\mathcal{H}_2$-norm for bilinear systems and discuss its meaning and usefulness for edge selection; there we also present our main theoretical result, showing the supermodularity of the $\mathcal{H}_2$-norm of bilinear systems when increasing the number of attacked edges; 
in Section \ref{sec:optimization}, we show how knowledge of the $\mathcal{H}_2$-norm of the system can help to target the protection of vulnerable edges, and propose three algorithms to estimate the best set of edges to protect as well as a way to compute a lower bound for the optimal solution, so we can evaluate the relative results of each algorithm; in Section \ref{sec:simul}, we simulate three different network topologies: a ring graph, a Barab\'asi-Albert graph and a simplified flight network graph, showing how $\mathcal{H}_2$-based selection algorithms perform when compared to a lower bound for the optimal solution;  finally, in Section \ref{sec:conclusions}, we conclude our work and make brief remarks about other potential problems where our main result may also be useful. 

\section{Preliminary Definitions}
\label{sec:prelim}

\subsection{Notations and Assumptions}
\label{sub:Notations}

Throughout this paper, the set of real numbers, non-negative real numbers, and strictly positive real numbers are represented as $\mathbb{R}$, $\mathbb{R}_+$, and $\mathbb{R}_{++}$, respectively. Similarly, the set of the strictly positive integers is denoted by $\mathbb{N}$ and the set of strictly positive integers up to $m$ by $\mathbb{N}_{\leq m}$. 
Matrices are represented by uppercase letters, and for a given matrix $N_k$, its element in row $i$ and column $j$ is represented as $n_k^{ij}$. Furthermore, we use the notation $N_k = \left(n_k(i, j)\right)_{ij}$ to bring attention to the expression of the individual elements of the matrix $N_k$ instead of the matrix itself. For any square matrix $M\in\mathbb{R}^{n\times n}$ let the operator $\tr(\cdot):\mathbb{R}^{n\times n} \rightarrow \mathbb{R}$ be $\sum_{k=1}^{n}m^{kk}$, the sum of the elements of the main diagonal of $M$. For any finite set of elements $\mathcal{V}$, let $|\mathcal{V}|\in\mathbb{N}\cup\{0\}$ be the number of elements in the set with $|\mathcal{V}|=0\iff\mathcal{V}=\emptyset$, and let $2^\mathcal{V}$ be the power set of $\mathcal{V}$. Furthermore, for any function $f$ with domain in $\mathcal{V}$, for each subset $\bar{\mathcal{V}}\subset\mathcal{V}$ define $f(\bar{\mathcal{V}}) = \{f(v) \ | \ v\in\bar{\mathcal{V}}\}$. The elementary vector of index $i$ is denoted $e_i$ and is a vector of all elements zero except for the $i$-th element, which is $1$. Similarly, an elementary matrix $E_{ij}=e_ie_j^\top$ has all its elements zero except for the one in position $(i,j)$, which is one.

 

\subsection{Bilinear Dynamical Networks}

    Consider a linear digraph given by the triplet $\mathcal{G}=(\mathcal{V},~\mathcal{E},~w)$. If $\mathcal{V}\subset\mathbb{N}$ is a set of agents with one-dimensional linear dynamics, $\mathcal{E}\subset \mathcal{V}\times\mathcal{V}$ is a set of edges, or interconnections between the agents, and $w:\mathcal{E}\rightarrow\mathbb{R}$ is the edge weight function (not assumed to be restricted in sign), then we call $\mathcal{G}$ a \emph{linear dynamical network}. We assume that the agent dynamics is given by
    \begin{equation}
        \label{eq:AgtDyn}
        \Sigma_i: ~\dot{x}_i(t) = -d_ix_i(t)+\sum_{j\rightarrow i}w(j,i)x_j(t),
    \end{equation}
    where $d_i\in\mathbb{R}$. More complicated linear dynamics for the individual nodes can be analyzed in the same manner as we do in this paper, but each edge must always connect the outputs of an agent to the state dynamics of the other linearly.
    
    Composing the states of the network into a vector $x = [x_1, ~x_2, ~\dots, ~x_n]^\top$, where $n=|\mathcal{V}|$ is the number of nodes in the network, we can write the dynamics of the complete network as
    \begin{equation}
        \label{eq:AutNetDyn}
        \Sigma: ~\dot{x}(t) = \underbrace{(-D+A^\top)}_{N_0}x(t),
    \end{equation}
    \noindent where $D=\diag(d_1,~d_2,~\dots,~d_n)$ and $A$ is the adjacency matrix of the graph, that is, $a^{ij}=w(i,j)$ if is there is an edge from node $i$ to node $j$, and $0$ otherwise. We then introduce the following definitions:
    
    \begin{definition}[Additive Disturbance]
        An \emph{additive disturbance} at node $i$ is a function $v_i:\mathbb{R}^+\rightarrow\mathbb{R}$ that alters the node dynamics as follows:
        \begin{equation}
            \label{eq:AddDstAgtDyn}
            \Sigma_i: ~\dot{x}_i(t) = -d_ix_i(t)+v_i(t)+\sum_{j\rightarrow i}w(j,i)x_j(t).
`        \end{equation}
        If a linear network $\mathcal{G}$ is subject to additive disturbances on a subset $\mathcal{V}_a\subset\mathcal{V}$ of its nodes, with $|\mathcal{V}_a|=m_\text{v}$, the dynamics of the network can be written as
        \begin{equation}
            \label{eq:AddDstAutNetDyn}
            \Sigma: ~\dot{x}(t) = N_0x(t)+Bv(t),
        \end{equation}
        \noindent where $v$ is the concatenation of the signals $v_i$ for every $i\in\mathcal{V}_a$, and $B\in\mathbb{R}^{n\times m_\text{v}}$ is the column composition of elementary vectors $e_i\in\mathbb{R}^n$, for every $i\in\mathcal{V}_a$ and in the same order as in $v$. 
        
        From now on we assume given an arbitrary order on the set $\mathcal{V}_a$, so $v_i$ and $b_i$ will refer respectively to the $i$-th element of $v$ and $i$-th column of $B$, which refer to the $i$-th node in set $\mathcal{V}_a$, and not necessarily to node $i$.
    \end{definition}
    
    \begin{definition}[Multiplicative Disturbance]
        A \emph{multiplicative disturbance} at edge $(i,j)$ is a function $\eta_{ji}:\mathbb{R}^+\rightarrow\mathbb{R}$ that alters the node dynamics as follows:
        \begin{equation}
            \label{eq:MulDstAgtDyn}
            \Sigma_i: ~\dot{x}_i(t) = -d_ix_i(t)+\sum_{j\rightarrow i}(w(j,i)+\eta_{ji}(t))x_j(t).
        \end{equation}
        If a linear network $\mathcal{G}$ is subject to multiplicative disturbances in a set $\mathcal{E}_a\subset\mathcal{E}$ of its edges, with $|\mathcal{E}_a|=m_\text{e}$, the dynamics of the network can be written as
        \begin{equation}
            \label{eq:MulDstAutNetDyn1}
            \dot{x}(t) = N_0x(t)+\sum_{\forall (i,j)\in\mathcal{E}_a}E_{ji}\eta_{ji}(t),
        \end{equation}
        \noindent where $E_{ji}=e_je_i^\top$ is called an elementary matrix. If we assume an arbitrary order for the elements of set $\mathcal{E}_a$ we can write \eqref{eq:MulDstAutNetDyn1} as
        \begin{equation}
            \label{eq:MulDstAutNetDyn2}
            \dot{x}(t) = N_0x(t)+\sum_{k=1}^{m_\text{e}}N_k\eta_{k}(t),
        \end{equation}
        \noindent where $N_k=E_{j_ki_k}$ and $\eta_k=\eta_{j_ki_k}$, $(i_k,j_k)$ being the $k$-th edge in $\mathcal{E}_a$. When convenient, we write the vector $\eta$ as the composition of $\eta_k$'s, $\eta = [\eta_1,~\eta_2,~\dots,~\eta_{m_\text{e}}]^\top$.
    \end{definition}
    
    \begin{remark}
        Notice that the case where the disturbances are undirected can be viewed as a particular case of the definition above, with $N_k=E_{ij}+E_{ji}$. As such, unless symmetry is explicitly assumed, any results obtained for directed graphs also holds for undirected graphs.
    \end{remark}
    
    Having defined additive and multiplicative disturbances we move to define a \emph{bilinear dynamical network} as follows:
    
    \begin{definition}[Bilinear Dynamical Network]
        A \emph{bilinear dynamical network} is a tuple $\mathcal{G}_{bl} = (\mathcal{V}, \mathcal{E}, w, \mathcal{V}_a, \mathcal{E}_a)$, where $\mathcal{G}=(\mathcal{V}, \mathcal{E}, w)$ describes a linear dynamical network, $\mathcal{V}_a\subseteq\mathcal{V}$ is an arbitrarily ordered set of $m_\text{v}$ attacked nodes, and $\mathcal{E}_a\subseteq\mathcal{E}$ is an arbitrarily ordered set of $m_\text{e}$ attacked edges.
        
        A bilinear dynamical network has dynamics $\Sigma$ as below:
        \begin{equation}
            \label{eq:BlnNetDyn}
            \Sigma:
                \dot{x}(t) = \left (N_0 + \sum\limits_{k = 1}^{m_\text{e}}\eta_k(t)N_k\right )x(t) + Bv(t)
        \end{equation}
        \noindent where $N_0\in\mathbb{R}^{n\times n}$ is defined as in \eqref{eq:AutNetDyn}, $B\in\mathbb{R}^{n\times m_\text{v}}$ is defined as in \eqref{eq:AddDstAutNetDyn}, $v$ is a vector of additive disturbances, $N_k\in\mathbb{R}^{n\times n}$ are defined as in \eqref{eq:MulDstAutNetDyn2}, and the $\eta_k$'s are multiplicative disturbances.
    \end{definition}
    
    We notice that the dynamics of a bilinear dynamical network is a particular case of the generic bilinear system given by
    \begin{equation}
        \label{eq:bilinBbar}
        \begin{cases}
            \dot{x}(t) = \left (N_0 + \sum\limits_{k = 1}^{m}u_k(t)\bar{N}_k\right)x(t) + \bar{B}u(t) \\
            y(t) = Cx(t)
        \end{cases}
    \end{equation}
    \noindent where $m=m_\text{v}+m_\text{e}$, $u=[\eta^\top, v^\top]^\top$, $\bar{N}_k = N_k$ for $1\leq k \leq m_\text{e}$ and $\bar{N}_k=0_{n\times n}$ for $k>m_\text{e}$, $\bar{B} = [0_{n\times m_\text{e}}, B]$, and $C=I_{n\times n}$. As a particular case, any results from the bilinear systems literature are immediately applicable to bilinear dynamical networks.

\subsection{The Volterra Series and the Solution of Bilinear Systems}

\begin{figure}
    \centering
    \includegraphics[scale = 0.8]{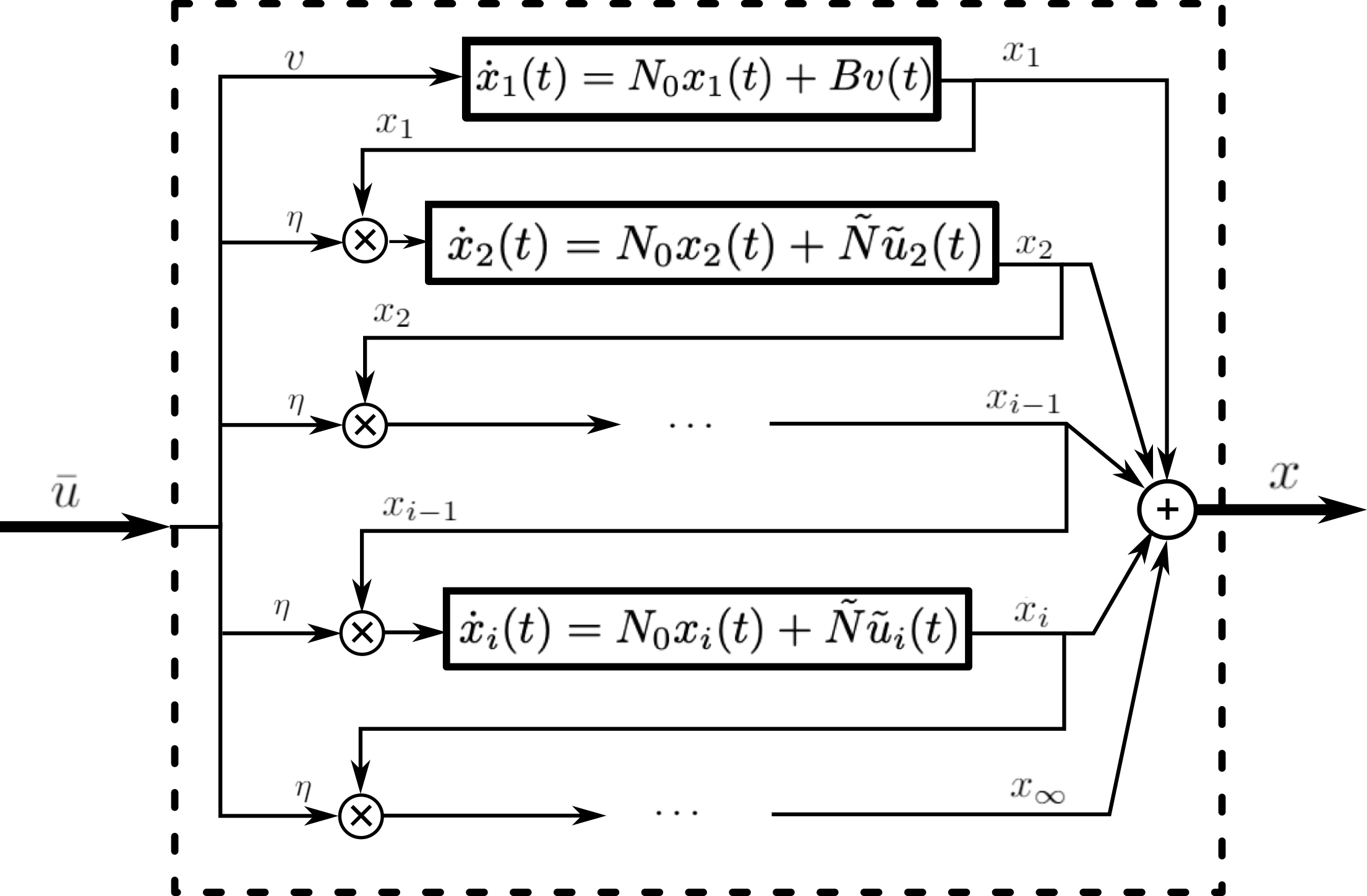}
    \caption{Graphical representation the bilinear system written  as the sum of interconnected linear systems.}
    \label{fig:BilinearSumOfLinear}
\end{figure}

To study the behaviours of bilinear dynamical networks, we first look at their solutions. To do that, notice that \eqref{eq:BlnNetDyn} can be rewritten as an infinite sum of the solutions of interconnected linear systems as follows:
\begin{equation}
    \label{eq:BilinearSumOfLinear}
    \begin{array}{l}
    \begin{cases}
        \dot{x}_1(t) = N_0x_1(t)+Bv(t)  \\
         \dot{x}_2(t) = N_0x_2(t)+\sum_{k=1}^mN_kx_{1}(t)\eta_k(t) \\
        ~~~~~~~~~~~~\vdots \\
         \dot{x}_i(t) = N_0x_i(t)+\sum_{k=1}^mN_kx_{i-1}(t)\eta_k(t) \\
        ~~~~~~~~~~~~\vdots \\
        \end{cases} \\
    \end{array}
\end{equation}
where $x(t) = \sum_{i=1}^{\infty}x_i(t)$.
Fig. \ref{fig:BilinearSumOfLinear} shows a graphical representation of this. If this infinite series is uniformly convergent, formally we have:
\begin{equation}
    \sum_{i=1}^\infty\dot{x}_i = N_0\sum_{i=1}^\infty x_i+\sum_{k=1}^mN_k\eta_k\sum_{i=1}^\infty x_i+Bv.
\end{equation}

By writing $\tilde{N} = [N_1, \dots, N_m]$ and $\tilde{u}_i(t) = [\eta_1x_{i-1} \ ; \ \dots \ ; \ \eta_{m_\text{e}}x_{i-1}]$ we can rewrite the complete systems in this form

\begin{equation}
    \begin{array}{l}
       \dot{x}_1(t) = N_0x_1(t)+Bv(t)  \\
       \dot{x}_2(t)= N_0x_2(t)+\tilde{N}\tilde{u}_{2}(t)\\
        ~~~~~~~~\cdots \\
        \dot{x}_i(t) = N_0x_i(t)+\tilde{N}\tilde{u}_{i}(t)\\
        ~~~~~~~~\cdots 
    \end{array}
\end{equation}

\noindent which, assuming zero initial condition, has solutions as shown below
\begin{equation}
    \begin{array}{l}
         x_1(t) = \int_0^t\mbox{e}^{N_0(t-\tau_1)}Bv(\tau_1)d\tau_1 \\
         x_2(t) = \int_0^t\mbox{e}^{N_0(t-\tau_2)}\tilde{N}\tilde{u}_2(\tau_2)d\tau_2\\
        ~~~~~~~~\cdots \\
         x_i(t) = \int_0^t\mbox{e}^{N_0(t-\tau_i)}\tilde{N}\tilde{u}_i(\tau_i)d\tau_i\\
        ~~~~~~~~\cdots 
    \end{array}
\end{equation}
After substituting the expression for $\tilde{N}$ and $\tilde{u}_i(t)$ in the solutions of the linear systems we get
\begin{equation}
    \begin{array}{l}
         x_1(t) = \int_0^t\mbox{e}^{N_0(t-\tau_1)}Bu(\tau_1)d\tau_1 \\
         x_2(t) = \int_0^t\int_0^{\tau_2}\mbox{e}^{N_0(t-\tau_2)}\sum_{k_2=1}^mN_{k_2}\eta_{k_2}\mbox{e}^{N_0(\tau_2-\tau_1)}Bvd\tau_1d\tau_2\\
         ~\cdots \\
         x_i(t) = \int_0^t\int_0^{\tau_i}\cdots\int_0^{\tau_2}\mbox{e}^{N_0(t-\tau_i)}\sum_{k_i=1}^mN_{k_i}\eta_{k_i}\mbox{e}^{N_0(\tau_i-\tau_{i-1})}\\ \times\sum_{k_i=1}^mN_{k_{i-1}}\eta_{k_{i-1}}\dots \mbox{e}^{N_0(t-\tau_2)}\sum_{k_2=1}^mN_{k_2}\eta_{k_2}\mbox{e}^{N_0(\tau_2-\tau_1)}\\ \times Bvd\tau_1d\tau_2\dots d\tau_i.
    \end{array}
\end{equation}
Adding all terms $x_i(t)$ together results in the Volterra series of the system presented in equation \eqref{eq:VolterraSeriesState}


\begin{equation}
    \label{eq:VolterraSeriesState}
    \begin{split}
        x(t) = &\sum_{i=1}^\infty\int_0^t\int_0^{\tau_i}\cdots\int_0^{\tau_2}\sum_{k_2,\dots,k_i=1}^{m}\mbox{e}^{N_0(t-\tau_i)}N_{k_i}\mbox{e}^{N_0(\tau_i-\tau_{i-1})}\\ &\times N_{k_{i-1}}\mbox{e}^{N_0(\tau_{i-1}-\tau_{i-2})}\cdots N_{k_2}\mbox{e}^{N_0(\tau_{2}-\tau_1)}B\\ &\times \eta_{k_i}(\tau_i) \eta_{k_{i-1}}(\tau_{i-1})\cdots \eta_{k_2}(\tau_2)v(\tau_1) d\tau_1d\tau_2\cdots d\tau_i.
    \end{split}
\end{equation}



\noindent where we represent the iterated sums $\sum_{k_1}\sum_{k_2}\dots\sum_{k_n}$ as $\sum_{k_1,k_2,\dots k_n}$. With this we can state the following definition.

\begin{definition}
    The Volterra kernels of the bilinear systems can be defined as

    \begin{equation}
        \label{eq:kernels}
        \begin{split}
            h_i&(t,\tau_1,\dots\tau_i) = \sum_{k_2,\dots,k_i=1}^{m}\mbox{e}^{N_0(t-\tau_i)}N_{k_i}\mbox{e}^{N_0(\tau_i-\tau_{i-1})}\\ &\times N_{k_{i-1}}\mbox{e}^{N_0(\tau_{i-1}-\tau_{i-2})}\cdots N_{k_2}\mbox{e}^{N_0(\tau_{2}-\tau_1)}B
        \end{split}
    \end{equation}
    
    \noindent and their multivariable Laplace transforms are $H_i(s_1, \dots, s_i) = \mathcal{L}(h_i(t_1,\dots,t_i))$, also called the $i$-th order transfer functions of the bilinear system.
\end{definition}

\begin{remark}
    Notice that the Volterra Kernels can have multiple equivalent definitions (see \cite{varona2018impulse}), and thus, this definition is not unique.
\end{remark}

Since we are using an infinite sum of dynamical systems, special care is needed when analysing convergence. The study of volterra series for general nonlinear systems is well explored in the literature, from which we can draw the following results:

\begin{definition}
    For a Volterra series with kernels $h_i$ given by \eqref{eq:kernels}, we define its \emph{gain bound function} for $x\geq 0$ as $f(x):=\sum_{i=1}^{\infty}\|h_i\|_{\infty}x^i$, and its \emph{radius of convergence} as $\rho := (\lim\sup_{n\rightarrow\infty}\|h_i\|_\infty^{1/n})^{-1}$.
\end{definition}

\begin{theorem}[Gain Bound Theorem \cite{boyd1984analytical}]
    \label{thm:GBT}
    A Volterra series \eqref{eq:VolterraSeriesState} with kernels $h_i$, gain bound function $f$ and radius of convergence $\rho$ converges absolutely for inputs with $\|\bar{u}\|_\infty<\rho$
\end{theorem}

From Theorem \ref{thm:GBT} we can conclude that for sufficiently bounded inputs, for any finite $T>0$ and a stable $N_0$, the bilinear system in equation \eqref{eq:bilinBbar} has a well defined solution given by \eqref{eq:VolterraSeriesState}. 

While the previous theorem assures the existence of a solution, we can also check its stability. Consider the definition below:

 \begin{definition}[\cite{sontag1998comments}]
    A dynamical system is \emph{integral input to state stable} (iISS) if there exist $\alpha, ~\gamma\in\mathcal{K}_\infty$ and $\beta\in\mathcal{KL}$ so that for all initial states $\xi$ and inputs $u(\cdot)$, its solution $x(t)$ respects
    
    \begin{equation}
        \alpha(|x(t)|)\leq\beta(|\xi|,t)+\int_0^t\gamma(|u(s)|)ds
    \end{equation}
    
    \noindent for all $t\geq0$.
\end{definition}

Then we can state the following theorem:

\begin{theorem}[\cite{sontag1998comments}]
    The bilinear system \eqref{eq:bilinBbar} is iISS if and only if $N_0$ is Hurwitz.
\end{theorem}

Notice that iISS, while not as strong as ISS, still means that any input with ``finite energy'' (as measured by $\gamma$) cannot make the system unstable.

\section{$\mathcal H_2$-Based Performance Measure and Its Properties}
\label{sec:theory}

In this section, we define and explore the meaning of the $\mathcal{H}_2$-norm as a vulnerability metric for the system given by equation \eqref{eq:BlnNetDyn} when subject to attacks on its links. For linear systems, it is shown that the $\mathcal{H}_2$-norm is an effective robustness metric \cite{siami2016fundamental, Bamieh12, Jadbabaie13}, which makes it appropriate for measuring vulnerabilities to external attacks.  We investigate some properties of the $\mathcal{H}_2$-norm for a class of bilinear systems.
 It is shown that the $\mathcal H_2$-norm of these systems can be interpreted as the convolution kernel energy (a.k.a.\ Volterra kernel energy) \cite{Benner, zhang2002h2}.

\subsection{$\mathcal{H}_2$-norm of bilinear systems}




In bilinear systems literature, the $\mathcal{H}_2$-norm is defined as a function of the multivariable Laplace transform of the volterra kernels as below.

\begin{definition}
    \label{def:H2normdef}
    Assuming zero initial condition and that the system is iISS, and letting the $i$-th order transfer function $H_i(s_1,s_2,\dots,s_i)$ be the multivariable Laplace transform of the $i$-th kernel $h_i(t_1,t_2,\dots,t_i)$, the $\mathcal{H}_2$-norm of a bilinear system $\Sigma$ is defined as
    \begin{equation}
        \label{eq:H2normdef}
        \begin{split}
            \|\Sigma\|_{\mathcal{H}_2} = & \Bigg(\sum_{i=1}^{\infty}\int_{-\infty}^{\infty}\cdots\int_{-\infty}^{\infty} \tr(H_i^\top(jw_1,\dots,jw_i)\\& \times H_i(jw_1,\dots,jw_i))dw_1\dots dw_i\Bigg)^{1/2}
        \end{split}
    \end{equation}
    %
    
    % ...
    
    
    
    
\end{definition}


%
%
With this definition, we can state the following theorem from \cite{zhang2002h2} to quantify the value of the $\mathcal H_2$ norm of bilinear system \eqref{eq:bilinBbar}.

\begin{theorem}[\cite{zhang2002h2} ]
    \label{thm:gramian}
Suppose the Volterra series \eqref{eq:VolterraSeriesState} uniformly converges on the interval $[0,\infty)$, then
    the input-state $\mathcal H_2$ norm of bilinear system \eqref{eq:BlnNetDyn} can be computed by
    \begin{equation}
    \label{eq:h2norm}
    \|\Sigma\|_2 = \sqrt{\tr(P)},
\end{equation}
where $P$ is the reachability Gramian of the bilinear system defined as 
\begin{equation}
    \label{eq:gramiandef}
    P = \sum_{q=1}^\infty\int_0^\infty\cdots\int_0^\infty 
    ~P_qP_q^\top ~dt_1 \cdots dt_q,
\end{equation}
where
\begin{equation}
    \label{eq:PqPqTGen}
    P_qP_q^\top = \mbox{e}^{N_0t_q}\sum_{k \in\mathcal{E}_a}N_kP_{q-1}P_{q-1}^\top N_k^\top \mbox{e}^{N_0^\top t_q},
\end{equation}
for $q>1$, and
\begin{equation}
    \label{eq:P1P1T}
    P_1P_1^\top = \mbox{e}^{N_0t_1}BB^\top \mbox{e}^{N_0^\top t_1}.
\end{equation}

Furthermore, if \eqref{eq:gramiandef} converges then $P$ is the solution of the generalized Lyapunov equation
\begin{equation}
    \label{eq:genlyapunov}
    N_0P + PN_0^\top + \sum_{k \in\mathcal{E}_a}N_kPN_k^\top + BB^\top = 0,   
\end{equation}
and is positive definite.

\end{theorem}

Theorem \ref{thm:gramian} allows the efficient computation of the $\mathcal{H}_2$-norm, enabling its use in optimization problems. These results, however, assume the convergence of the infinite sums in \eqref{eq:H2normdef} and \eqref{eq:gramiandef}. In \cite{zhang2002h2} the authors state the following assumption as sufficient for that:

\begin{assumption}
\label{assump:1}
    The matrix $N_0$ is stable and for two numbers $\alpha$ and $\beta$, which satisfy the inequality $\|\mbox{e}^{N_0t}\| \leq \beta \mbox{e}^{-\alpha t}$ for all $t>0$, we have $\sqrt{\sum_{k \in\mathcal{E}_a}\|N_kN_k^\top\|}<\sqrt{2\alpha} / \beta$.
\end{assumption}

Interpreting the assumption we notice that we are asking for the linear dynamics to be dominant over the worst possible case of the bilinear dynamics. As we will verify in the simulations section, this assumption is often too restrictive and later on we will explore an alternative, more relaxed assumption that has originates from stochastic bilinear network literature and has been recently shown in \cite{redmann2019bilinear} to hold meaning for the deterministic case.

\subsection{The $\mathcal{H}_2$-norm as a Performance Metric for Bilinear Systems}

The $\mathcal{H}_2$-norm is a popular robustness metric for linear systems analysis (see \cite{siami2016tractable,siami2013fundamental,morris2015using}) since it is directly proportional to the expected deviation of the norm of the output when subject to a white noise disturbance. For nonlinear systems, however, the concept of the $\mathcal{H}_2$-norm is difficult to generalize.

Consider the bilinear system written as in equation \eqref{eq:bilinBbar} and assume that all inputs $\bar{u}$ are independently sampled white noise signals $\bar{u} = \eta = dW/dt$, with unitary covariance. The resulting stochastic differential equation (SDE) can be written as

\begin{equation}
    \label{eq:bilinSDE}
    dx(t) = N_0x(t)dt+\sum_{k=1}^{m}(N_kx(t)+\bar{b}_k)dW_k(t),
\end{equation}

\noindent where the $W_k$'s are independently sampled Brownian motion signals.

By taking the expected value of the integral form of equation \eqref{eq:bilinSDE}, we can find the dynamic equations of the average of the state $\bar{x} = \mathbb{E}(x)$ as

\begin{equation}
    \label{eq:meanEq}
    \dot{\bar{x}}(t) = N_0\bar{x}(t),
\end{equation}

\noindent which depends only on the drift matrix. This equation is another way to show that $N_0$ being Hurwitz is enough for stability of the system subject to white noise inputs. Furthermore, from the literature \cite{Tobias2004,priestley1988non,damm2004rational} we collect the following result for bilinear system subject to white noise inputs

\begin{assumption}
    \label{assump:2}
    Either for a stochastic bilinear system as in \eqref{eq:bilinSDE} or for a deterministic bilinear system as in \eqref{eq:bilinBbar}, $N_0$ is Hurwitz and the following holds:
    
    \begin{equation}
        \sigma_{\max}\left(I\otimes N_0+N_0\otimes I + \sum_{k=1}^{m}N_k\otimes N_k\right)<0
    \end{equation}
\end{assumption}

\begin{theorem}
    For a bilinear SDE as in equation \eqref{eq:bilinSDE} that respects Assumption \ref{assump:2}, we can say that
    
    
    \begin{equation}
        \lim_{t\rightarrow\infty}\mathbb{E}(x(t)) = 0,
    \end{equation}
    
    \noindent and
    
    \begin{equation}
        \lim_{t\rightarrow\infty}\cov(x(t)) = P,
    \end{equation}
    
    \noindent where $P$ is the solution of the generalized Lyapunov Equation
    
    \begin{equation}
        N_0P+PN_0^\top+\sum_{k=1}^{m}(N_kPN_k^\top+\bar{b}_k\bar{b}_k^\top)=0
    \end{equation}
    
\end{theorem}
\begin{proof}
    We start by taking the expected value of the integral form of \eqref{eq:bilinSDE}, as done in \cite{damm2004rational}, resulting in \eqref{eq:meanEq} which implies the first statement of the theorem if $N_0$ is Hurwitz. For the covariance, since as time goes to infinity the average of the state goes to zero, it converges to the second momentum, that is
    
    \begin{equation}
        \lim_{t\rightarrow\infty}\cov(x) = \lim_{t\rightarrow\infty}\mathbb{E}((x-\bar{x})(x-\bar{x})^\top) = \lim_{t\rightarrow\infty}\mathbb{E}(xx^\top).
    \end{equation}
    
    We compute the derivative of the second momentum $\mathbb{E}(xx^\top)$ using Ito's lemma from
    
    \begin{equation}
        \begin{split}
            d(xx^\top) = &x\bigg(x^\top N_0^\top dt+\sum_{k=1}^{m}(x^\top N_k^\top + \bar{b}_k^\top)dW_k\bigg)+\\&+\bigg(N_0xdt+\sum_{k=1}^{m}(N_kx+\bar{b}_k)dW_k\bigg)x^\top+\\&+\Bigg(\sum_{k_1=1}^{m}(N_{k_1}x+\bar{b}_{k_1})dW_{k_1}\Bigg)\\&\times \Bigg(\sum_{k_2=1}^{m}dW_{k_2}(x^\top N_{k_2}^\top + \bar{b}_{k_2}^\top)\Bigg).
    \end{split}
    \end{equation}
    
    \noindent which can be simplified by taking the expected value to 
    
    \begin{equation}
        \label{eq:pdot}
        \begin{split}
            &\dot{P} = PN_0^\top+N_0P+\\&+\sum_{k=1}^{m}(N_{k}PN_{k}^\top + N_{k}\mathbb{E}(x)\bar{b}_{k}^\top+\bar{b}_{k}\mathbb{E}(x)^\top N_{k}^\top+\bar{b}_{k}\bar{b}_{k}^\top)
        \end{split}
    \end{equation}
    
    \noindent since $\mathbb{E}(dW_{k_1}dW_{k_2})=0$ if $k_1\neq k_2$ and $\mathbb{E}(dW_{k}dW_{k})=dt$. 
    
    We can easily verify that the ODE is stable as long as Assumption \ref{assump:2} holds. Then, as time goes to infinity we get
    
    \begin{equation}
        0 = PN_0^\top+N_0P+\sum_{k=1}^{m}N_{k}PN_{k}^\top + \bar{B}\bar{B}^\top
    \end{equation}
    
    \noindent which completes the proof.
    
\end{proof}
    \begin{remark}
        We can show that Assumption \ref{assump:1} implies Assumption \ref{assump:2} and is, therefore, sufficient for stability of the solution of the equation above. Take the $L_2$-norm as in
    
    \begin{equation}
        \begin{split}
            \|\dot{P}\| &= \|N_0P+PN_0^\top+\sum_kN_kPN_k^\top+BB^\top\|\leq \\ &\leq 2\|N_0P\|+\|\sum_kN_kPN_k^\top\|+\|BB^\top\|\leq \\
            &\leq 2\sigma_{max}(N_0)\|P\|+\|\sum_kN_kN_k^\top\|\|P\|+\|BB^\top\|,
        \end{split}
    \end{equation}
    
    \noindent which is a monovariable deterministic ODE with stable solution if and only if $\|\sum N_kN_k^\top\|\leq 2\|\sigma_{max}(N_0)\|$. Therefore, if that upper bound on the norm of $P$ is stable, then $P$ converges.
    
    \end{remark}

    
    
    
    

\begin{theorem}[\cite{redmann2019bilinear}]
    Let $y$ be the output of system \eqref{eq:bilinBbar} with $x_0=0$ and suppose that Assumption \ref{assump:2} holds. Then, we can write that
    
    \begin{equation}
        \label{eq:redmanresult}
        \sup_{t\geq 0}\|y(t)\|_2 \leq \left(\trace(CPC^\top)\right)^{\frac{1}{2}}\exp\left\{0.5\|u^0\|^2_{\mathcal{L}^2}\right\}\|u\|_{\mathcal{L}^2},
    \end{equation}
    where $P$ is the unique solution to
    
    \begin{equation}
        \label{eq:redmanlyapeq}
        N_0P+PN_0^\top + \sum_{k=1}^m\bar{N}_kP\bar{N}_k^\top = -\bar{B}\bar{B}^\top.
    \end{equation}
\end{theorem}

The vector $u^0$ is defined in \cite{redmann2019bilinear} as $u^0_k(t) = u_k(t)$ if $N_k\neq 0$ and $u^0_k(t)=0$ otherwise, that is $u^0$ is nonzero only on the inputs that affect the system in a bilinear way. The first thing to notice from the result above is that it recovers the linear definition of the $\mathcal{H}_2$-norm, that is if $N_k=0$ for all $k$ then \eqref{eq:redmanlyapeq} reduces to the linear Lyapunov Equation and \eqref{eq:redmanresult} can be simplified to

\begin{equation}
    \trace(CPC^\top) = \|\Sigma\|_{\mathcal{H}_2}^2 = \sup_{u\in L_2}\frac{\|y\|_{L_\infty}}{\|u\|_{L_2}}.
\end{equation}

For bilinear system, the result still establishes a $L_2$ to $L_\infty$ relationship for the system, although it is not as direct as an induced norm. In fact, a remark in \cite{redmann2019bilinear} shows that as long as $N_0$ is Hurwitz, one can restrict the inputs to a small enough ball around $0$ such that \eqref{assump:2} holds, which is exactly the same condition proven in \cite{sontag1998comments} as necessary and sufficient for a bilinear system to be iISS, a established condition for $L_2$ to $L_\infty$ stability of nonlinear systems.

So from stochastic bilinear system literature we have a direct relationship between the $\mathcal{H}_2$-norm and the covariance of the output subject to white noise inputs. From the deterministic bilinear systems literature we have recent results showing that the bilinear $\mathcal{H}_2$-norm is a measure of the relation between the $L_2$-norm of the inputs and the $L_{\infty}$-norm of the output. This reinforces and consolidates the value of the $\mathcal{H}_2$-norm as a performance metric for bilinear system. In the next section we seek to characterize the $\mathcal{H}_2$-norm as a set function on the set of vulnerable edges and look for useful properties that can help efficiently solve the edge selection problem.

\subsection{Supermodularity of the $\mathcal H_2$ Norm}

Our objective in this section is to characterize the $\mathcal{H}_2$-norm of the bilinear system as a function of the set of edges under attack $\mathcal E_a$. For the main theoretical result of this paper, consider the following definition

\begin{definition}
    Define a \emph{family of bilinear digraphs} $\mathcal{F}$ generated by the ground set of $m_v$ vulnerable edges $\mathcal{E}_v\subseteq \mathcal{V}\times\mathcal{V}$ as follows:
    \[ \mathcal F(\mathcal{E}_v):= \left \{\mathcal{G} = (\mathcal{V}, \mathcal{E}, w, \mathcal{E}_a, \mathcal{V}_a) \big |~ \forall \mathcal{E}_a\in 2^{\mathcal{E}_v} \right \},\]
    for given node set $\mathcal V$, edge set $\mathcal E$, weight function $w$, and attacked node set $\mathcal V_a$.
    We assume $\Sigma(\mathcal{E}_a)$ is the bilinear system \eqref{eq:BlnNetDyn} induced by the corresponding bilinear digraph $(\mathcal{V}, \mathcal{E}, w, \mathcal{E}_a, \mathcal{V}_a) \in \mathcal F(\mathcal E_v)$. We can, then, define the square of the $\mathcal H_2$ norm as a set function $\rho_{\Sigma}(.):2^{\mathcal{E}_v}\rightarrow \mathbb{R}_+$ as 
    \begin{equation}
        \rho_{\Sigma}(\mathcal{E}_a) ~:=~ \|\Sigma(\mathcal{E}_a)\|_{\mathcal{H}_2}^2,~~~\forall \mathcal{E}_a\in2^{\mathcal{E}_v}.
        \label{eq:rho}
    \end{equation}
\end{definition}

In the following theorem, we characterize some functional properties of set function $\rho_{\Sigma}(.): 2^{\mathcal{E}_v}\rightarrow \mathbb{R}_+$.
\begin{theorem}
    \label{thm:mainresult}
    Suppose that for a family of bilinear digraphs $\mathcal{F}$ the $\mathcal{H}_2$-norm is properly defined for everyone of its elements, then the square of the $\mathcal H_2$ norm defined as a set function $\rho_{\Sigma}(\mathcal{E}_a):2^{\mathcal{E}_v}\rightarrow \mathbb{R}_+$ is monotone and supermodular.

\end{theorem} 

The proof is presented at the appendix. 
    Notice that the assumption made on the theorem (proper definition of the $\mathcal{H}_2$-norm) is satisfied if all elements of $\mathcal{F}$ satisfy either Assumption \ref{assump:1} or \ref{assump:2}. To investigate the tightness and relationship between the two assumptions, and the behaviour of the system when they are broken, consider the following system:
    
    \begin{equation}
        \label{eq:ex_sys}
        \dot{x} = -ax+kx\eta+bv
    \end{equation}
    
    \noindent where $\eta$ and $v$ are independently sampled white noise inputs and $a$, $b$ and $k$ are positive nonzero constants.
    
    The generalized Lyapunov equation given by \eqref{eq:genlyapunov} can be solved for this system as
    \begin{equation}
            P = \frac{b^2}{2a-k^2},
    \end{equation}
    Assumption \ref{assump:1} can be written as $a>0$ and $|k|<\sqrt{2a}$, or $2a-k^2>0$, and Assumption \ref{assump:2} becomes $-2a+k^2<0$. Looking at the formulation for the Grammian from equation \eqref{eq:gramiandef}, we can write 
    
    \begin{equation}
        \bar{P}_1 = \int_0^\infty e^{-a\tau}bbe^{-a\tau}d\tau = \frac{b^2}{2a}
    \end{equation}
    \begin{equation}
        \bar{P}_2 = \int_0^\infty e^{-a\tau}k\bar{P}_1ke^{-a\tau}d\tau = \frac{b^2}{2a}\frac{k^2}{2a}
    \end{equation}
    \begin{equation}
        \bar{P}_i = \int_0^\infty e^{-a\tau}k\bar{P}_{i-1}ke^{-a\tau}d\tau = \frac{b^2}{2a}\left(\frac{k^2}{2a}\right)^{i-1}
    \end{equation}
    
    \noindent with 
    
    \begin{equation}
        P = \sum_{i=1}^\infty\bar{P}_i = \sum_{i=1}^\infty\frac{b^2}{2a}\left(\frac{k^2}{2a}\right)^{i-1}.
    \end{equation}
    
    This defines the infinite sum of a geometric progression with quotient $q = k^2/2a$ and initial value $a_0 = b^2/2a$. The necessary and sufficient convergence condition for the sum is $k^2/2a<1 \iff 2a-k^2>0$ which coincides with Assumptions \ref{assump:1} and \ref{assump:2}. This means that for this SISO bilinear system, both assumptions coincide and are necessary and sufficient for any positive values of $k$, $a$ and $b$, instead of just the sufficiency for the general case.
    
    Running a quick simulation for such system with $a=1$, $b=1$ and different values of $k$ allows us to observe how breaking the convergence condition affects the behaviour of the system. 
    
    
    \begin{figure*}
         \centering
         \begin{subfigure}[b]{0.32\textwidth}
             \centering
             \includegraphics[scale = 0.43]{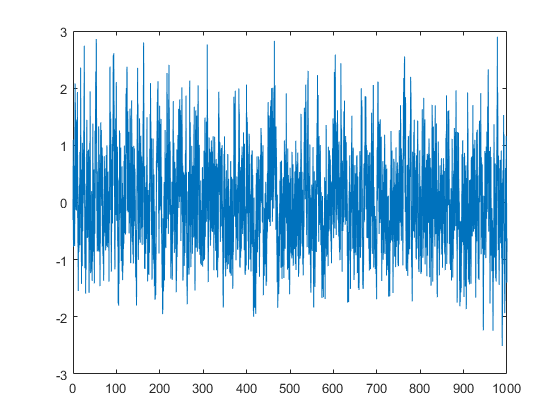}
             \caption{Simple Example Simulation for $k=0.1$.}
             \label{fig:SimpleSimk01}
         \end{subfigure}
        \hfill
         \begin{subfigure}[b]{0.32\textwidth}
             \centering
             \includegraphics[scale = 0.43]{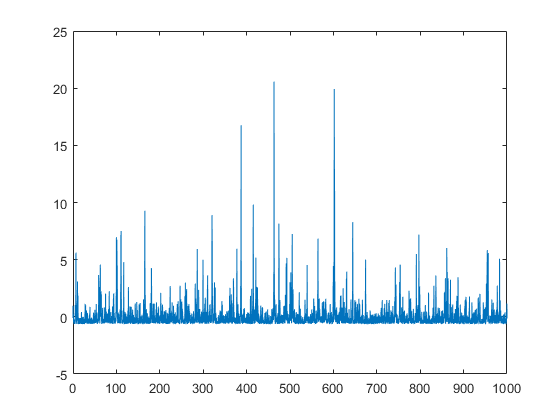}
            \caption{Simple Example Simulation for $k=\sqrt{2}$.}
            \label{fig:SimpleSimk14}
         \end{subfigure}
        \hfill
         \begin{subfigure}[b]{0.32\textwidth}
             \centering
             \includegraphics[scale = 0.43]{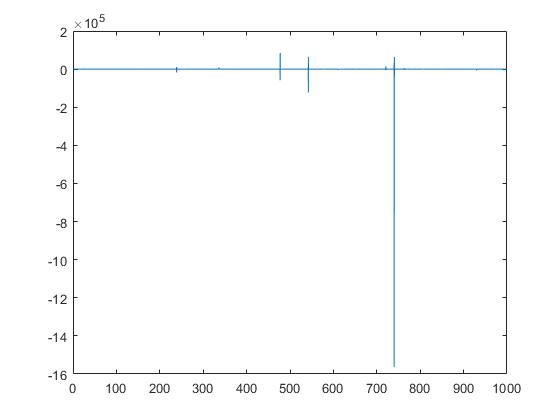}
            \caption{Simple Example Simulation for $k=10$.}
            \label{fig:SimpleSimk10}
         \end{subfigure}
            \caption{In this figure we show how varying the bilinear term $k$ affects the behaviour of a simple SISO bilinear system \eqref{eq:ex_sys}, where $\eta$ and $v$ are independently sampled white noise inputs and $a=1$, $b=1$ and $k$ is some positive nonzero constant.}
            \label{fig:SimpleSysSimul}
    \end{figure*}
    
    First consider Fig. $\ref{fig:SimpleSimk01}$, where $k=0.1$, clearly respecting both assumptions for this system. The simulation was done ten times and the averaged covariance is $0.5051$, very closely lower bounded by the Grammian $P = 0.5025$. We actually observed some of the samples with a covariance bellow $0.5025$, indicating the tightness of the bound for this system. 
    
    
    
    The second simulation presented in figure \ref{fig:SimpleSimk14} is of the system with $k=\sqrt{2}$, which is the borderline of assumption \ref{assump:1}. For this system, the estimate for the covariance of the state when subject to white noise input diverges, and we observe a change in the system's behaviour with high amplitude peaks. The experimental covariance of the system also varied in different runs of the simulation, but it always resulted in high, sporadic peaks. This is accentuated when $k=10$, as in Figure \ref{fig:SimpleSimk10}, where the peaks are more evident. 
    
    
    Notice that this system is iISS, so its deterministic response for $L_2$ inputs can never be unstable, no matter the value of $k$. The instability in the covariance of the state appears on the high peaks, which become more pronounced the bigger $k$ is in relation to $a$ and $b$.
    
\section{Optimal Strategy}
\label{sec:optimization}

After reviewing the tools present in the literature and identifying useful properties of the $\mathcal{H}_2$-norm in the context of edge selection, we turn our attention to the network synthesis problem. Out goal is to improve the performance of a bilinear network \eqref{eq:BlnNetDyn} by protecting and removing $\ell \geq 1$ edges from the vulnerable edge set $\mathcal E_v$. Specifically, we explore how to find a subset of $\ell$ vulnerable edges $\mathcal{E}_p\subset\mathcal{E}_v$, $|\mathcal{E}_p|\leq \ell$ that if protected minimizes the $\mathcal{H}_2$-norm of the system.

\subsection{Edge Protection Problem Formulation}

We start by pointing out the triviality of the problem from the point of view of the attacker. Since the $\mathcal{H}_2$-norm is supermodular under edge addition, for any linear cost functions that an attacker might need to balance, the best solution is to either attack as many edges as possible, or no edge at all. 
%

The optimization problem becomes more interesting under the perspective of the system designer, who wants to select the best specific set of edges to protect, subject to a limit on how many edges can actually be protected.


Assuming an attacker will always target as many edges as possible, we must protect enough edges to assure that either Assumption \ref{assump:1} or \ref{assump:2} holds. 
The edge protection problem can, then, be cast as the following combinatorial optimization problem
\begin{equation}
    \label{eq:optimincardinal}
    \begin{aligned}
        \mathcal{E}_* = \arg\min_{\mathcal{E}_a\subset\mathcal{E}_v} \quad & \rho_\Sigma(\mathcal{E}_a)\\
        \textrm{s.t.} \quad & |\mathcal{E}_a| \geq |\mathcal{E}_v|-\ell,
    \end{aligned}
\end{equation}
where $\mathcal E_v$ is the vulnerable edge set, and budget $\ell$ is the maximum number of edges we can protect. The optimal protected edge set can be obtained by 
 $\mathcal{E}_p = \mathcal{E}_v\backslash \mathcal{E}_*$ from \eqref{eq:optimincardinal}.

While combinatorial problems are usually hard to solve, we note that for $\ell=1$, one only needs to compute the value of $\rho_\Sigma(\mathcal{E}_a)$ for all the $n$ sets of attacked edges with $n-1$ elements to find the solution. In general, however, the number of sets with $n-\ell$ elements grows almost exponentially for values of $\ell$ close to $n/2$, i.e.,  ${n \choose \lceil n/2 \rceil} \sim \frac{2^{n}}{\sqrt{ n}}$. Therefore such a straightforward strategy could be computationally expensive.

\subsection{Linearized Edge Selection}

The first intuitive approach is to linearize the cost at some operation point. To do that, consider a relaxed cost function $\rho_\Sigma$ as

\begin{equation}
    \begin{split}
    \rho_\Sigma(c_k) &=\trace(P(c_k)) = \sum_{i=1}^n e_i^\top P(c_k)e_i = \\ &=\sum_{i=1}^n e_i^\top \otimes e_i^\top \mbox{vec}(P(c_k)) = \\
    &= \sum_{i=1}^n e_i^\top \otimes e_i^\top (A\otimes I + I\otimes A + \sum_{k=1}^{m_\text{e}}c_k^2N_k\otimes N_k)^{-1}\times \\ &\times\mbox{vec}(BB^\top),
    \end{split}
\end{equation}

\noindent where $c\in\mathbb{R}^n$, $c=[c_1, c_2, \dots, c_{m_\text{e}}]^\top$, with $0\leq c_k\leq 1$, for $k=1, 2, \dots, m_\text{e}$. It is easy to verify that such relaxed function is the extention of the set function $\rho_\Sigma$ to the polyhedron with vertices at the original domain set. Defining $W(c)$ as

\begin{equation}
    \label{eq:Wdef}
    W(c) = A\otimes I + I\otimes A + \sum_{k=1}^{m_\text{e}}c_k^2N_k\otimes N_k,
\end{equation}

\noindent we take the partial derivative with respect to $c_k$ at some operation point $\bar{c}$ as

\begin{equation}
\label{eq:partder}
\begin{split}
    \frac{\partial \rho_\Sigma}{\partial c_k}(\bar{c}) &= \sum_{i=1}^n(e_i^\top \otimes e_i^\top)\frac{\partial (W(\bar{c}))^{-1}}{\partial c_k}\mbox{vec}(BB^\top) = \\ &= \sum_{i=1}^n(e_i^\top \otimes e_i^\top)W(\bar{c})^{-1}\frac{\partial W}{\partial c_k}(\bar{c})W(\bar{c})^{-1}\mbox{vec}(BB^\top),
\end{split}
\end{equation}

\noindent where

\begin{equation}
    \frac{\partial W}{\partial c_k}(\bar{c}) = 2c_kN_k\otimes N_k.
\end{equation}

To select the $\ell$ edges to protect we evaluate \eqref{eq:partder} at $\bar{c} = [1, ~1, \dots, ~1]^\top$ for all $k=1,~ 2, ~\dots, ~m_\text{e}$, and select the $\ell$ highest values. The algorithm \ref{alg:linear} is used to implement this approach in the simulations.

\begin{algorithm}[t]
    \SetKwInOut{Input}{Input}
    \SetKwInOut{Output}{Output}

    \Input{$\Sigma$, $\mathcal E_v$, and $\ell$}
            \vspace{.1cm}
    \Output{$\mathcal E_*$}
    		\vspace{.4cm}
	$\mathcal E_a \leftarrow \{\}$\\	
	 \vspace{.1cm}
        \For{$k = 1$ {\it to} $m_{e}$}
       {$\partial\rho k(k)$  $\leftarrow$ compute $\frac{\partial\rho_\Sigma}{\partial c_k}$ for $\bar{c} = [1, ~1, ~\dots, ~1]^\top$\\
       }
       $i\leftarrow$ indexes of $\ell$ highest elements of $\rho k$\\
       $\mathcal E_*\leftarrow \mathcal{E}_v(i)$\\
       \vspace{.1cm}
        \Return $\mathcal E_*$
    \caption{\small An algorithm for a given bilinear system $\Sigma$ that select edges to protect by minimizing a linearized version of the cost $\rho_\Sigma$.}
    \label{alg:linear}
\end{algorithm}

\subsection{Greedy Edge Selection}

A second approach is to use a greedy algorithm by leveraging the fact that we can solve the problem for $\ell=1$ and that there are known theoretical bounds for the greedy minimization of supermodular functions subject to cardinality constraints. The greedy algorithm \ref{alg:greedyVarying} is used in the simulations.

 \begin{algorithm}[t]
    \SetKwInOut{Input}{Input}
    \SetKwInOut{Output}{Output}

    \Input{$\Sigma$, $\mathcal E_v$, and $\ell$}
            \vspace{.1cm}
    \Output{$\mathcal E_*$}
    		\vspace{.4cm}
	$\mathcal E_a \leftarrow \{\}$\\	
	 \vspace{.1cm}
        \For{$k = 1$ {\it to} $\ell$}
       { $\{e\}$ $\leftarrow$ find an edge in $\mathcal E_v$ that returns the minimum value for \[\rho_\Sigma\left( \mathcal E_a \cup \{e\}\right) - \rho_\Sigma (\mathcal E_a)\]\\
       $\mathcal E_a \leftarrow \mathcal E_a \cup \{e\}$\\
       $\mathcal E_v \leftarrow \mathcal E_v \backslash \{e\}$\\
       }
       $\mathcal E_*\leftarrow\mathcal E_a$\\
       \vspace{.1cm}
        \Return $\mathcal E_*$
    \caption{\small A greedy heuristic for a given bilinear system $\Sigma$ which sequentially picks edges to protect that minimize a  metric increase of $\rho_\Sigma$.}
    \label{alg:greedyVarying}
\end{algorithm}

It is known that for the maximization of submodular functions is NP-hard and the greedy algorithm does not deliver the optimal solution in general. However optimality gaps are given in the literature allowing the leverage of efficient algorithms to find a good suboptimal solution. 
In the following, we  present  a  key result from the theoretical computer science literature to efficiently obtain an approximate solution for this combinatorial optimization.

\begin{theorem}[\cite{buchbinder2014submodular}]\label{th:3}
    There exists an efficient algorithm with time complexity $\mathcal{O}(n\ell)$ that, given a non-negative submodular function $f$ and a cardinality parameter $\ell$, achieves an approximation greater than $0.356$ for the problem: $max\{f(\mathcal{S}):|\mathcal{S}|=\ell\}$, where $f(\cdot)$ is a submodular function.
\end{theorem}

In \cite{buchbinder2014submodular}, the authors present a greedy-based algorithm to prove Theorem \ref{th:3} and to guarantee  that the resulted solution is at least $35.6\%$ of the maximum value of the function. While it is reassuring to know the optimality gap is bounded, $35.6\%$ is a large margin for suboptimality. To evaluate a better margin, in the next section we relax the combinatorial optimization to obtain a closer lower bound for the optimal solution of this problem.


\subsection{A Lower Bound for the Optimal Cost}
\label{sub:lowbnd}

Next we compute a lower bound for the optimal solution of our problem. With this we can evaluate an upper bound for the optimality gap and check if the greedy solution is good.

To do that, first we relax the integer problem of selecting a subset $\mathcal{E}_a$ of a set of vulnerable edges $\mathcal{E}_v$, as in \eqref{eq:optimincardinal}, to looking for a vector $c\in\mathbb{R}^m$, where $m=|\mathcal{E}_v|$, $c=[c_1, c_2, \dots, c_m]^\top$, as below.

\begin{equation}
    \label{eq:rlxprb}
    \begin{aligned}
        \min_{P, c} \quad & \tr(P)\\
        \textrm{s.t.} \quad & N_0P+PN_0^\top+\sum_{k=1}^mc_k^2N_kPN_k+BB^\top=0, \\
        & \sum_{k}c_k \geq m_v-\ell, \\
        & 0\leq c_k \leq 1, ~\forall k=1,\dots,m.
    \end{aligned}
\end{equation}

Notice that the first constraint is quadratic and bilinear in the parameters. To solve the first problem we can redefine $\bar{c} = [c_1^2, ~c_2^2, \dots, c_m^2]^\top$. To solve the second problem, first notice that we can solve the generalized Lyapunov equation for $P$ by using the Kronecker product as

\begin{equation}
    \label{eq:solP}
    \mathbf{vec}(P) = -W(\bar{c})^{-1}\mathbf{vec}(BB^\top),
\end{equation}

\noindent where

\begin{equation}
    \label{eq:Wc}
    W(\bar{c}) = N_0\otimes I+I\otimes N_0 + \sum_k \bar{c}_k N_k\otimes N_k.
\end{equation}

Next, we point that for $N_0\in\mathbb{R}^{n\times n}$,

\begin{equation}
    \label{eq:tracelinear}
    \trace(N_0) = \sum_{k=1}^ne_i^\top N_0 e_i = \sum_{k=1}^n(e_i^\top\otimes e_i^\top)\mathbf{vec}(N_0),
\end{equation}

\noindent where $e_i$ is the $i$-th vector in the canonical base of the vector space. Using \eqref{eq:solP} and \eqref{eq:tracelinear}, we can rewrite the relaxed problem \eqref{eq:rlxprb} as

\begin{equation}
    \label{eq:lboptprob}
    \begin{aligned}
        \min_{\bar{c}} \quad & -\sum_{k=1}^n(e_k^\top\otimes e_k^\top) W(\bar{c})^{-1}\mathbf{vec}(BB^\top)  \\
        \textrm{s.t.} \quad & \sum_{k}\bar{c}_k \geq m_v-\ell, \\
        & 0\leq \bar{c}_k \leq 1, ~\forall k=0,\dots,m
    \end{aligned}
\end{equation}

{

The conversion of the constraint $\sum_kc_k\geq m_v-\ell$ to $\sum_kc_k^2\geq m_v-\ell$ does not result in the same feasibility set, but it is easy to verify that $\sum_kc_k^2\geq m_v-\ell\rightarrow \sum_kc_k\geq m_v-\ell$ if $0\leq c_k \leq 1$. As such a solution to \eqref{eq:lboptprob} is a lower bound to the solution of \eqref{eq:rlxprb}, which is a lower bound to the solution of the original combinatorial optimization problem. 

The final step for being able to efficiently solve our lower bound problem is to show that its cost function is convex. Consider the functions $f:\mathbb{R}^n\rightarrow\mathbb{R}$, $g:\mathcal{S}_{n^2}^+\rightarrow\mathbb{R}_+$ and $h:\mathbb{R}^n\rightarrow\mathcal{S}_{n^2}^+$ defined as below

\begin{equation}
    h(\bar{c}) = -W(\bar{c}) = -(N_0\otimes I+I\otimes N_0 + \sum_k \bar{c}_k N_k\otimes N_k),
\end{equation}
\begin{equation}
    \label{eq:gdef}
    g(W) = \sum_{k=1}^n(e_k^\top\otimes e_k^\top) W^{-1}\mathbf{vec}(BB^\top)
\end{equation}
\begin{equation}
    f(\bar{c}) = g(h(\bar{c})).
\end{equation}

One can easily verify that $f$ defined as above is the cost function of our relaxed problem, and that $h$ is affine on its arguments. One can also conclude that any $\bar{c}$ in the domain of definition of the $\mathcal{H}_2$ norm results in a positive definite value for $h(\bar{c})$ (since $W(\bar{c})$ needs to be negative definite for Assumption \ref{assump:1} to hold). For showing convexity of $g$ consider the following lemma:

\begin{lemma}
    Function $g(.)\,:\, \mathcal S_n^+ \rightarrow \mathbb R_+$ given by \eqref{eq:gdef} is convex.
\end{lemma}

\begin{proof}
    Function $g$ is convex if and only if 
    
    \begin{equation}
        g(\lambda W_1 + (1-\lambda)W_2) \leq \lambda g(W_1)+(1-\lambda)g(W_2).
    \end{equation}
    
    Defining $\bar{g}(W) = W^{-1}$ we rewrite the inequality above as
    
    \begin{equation}
        \begin{split}
            \sum_{k=1}^n(e_k^\top\otimes e_k^\top) \bar{g}(\lambda W_1+(1-\lambda)W_2)\mathbf{vec}(BB^\top) \leq \\ \sum_{k=1}^n(e_k^\top\otimes e_k^\top) (\lambda\bar{g}(W_1)+(1-\lambda)\bar{g}(W_2))\mathbf{vec}(BB^\top)
        \end{split}
    \end{equation}
    which allow us to conclude that $g$ is convex if $\bar{g}$ is convex in the positive definite sense. To show convexity of $\bar{g}$ we need to show that for two positive definite matrices $W_1$ and $W_2$, 
    
    \begin{equation}
        \lambda W_1^{-1} + (1-\lambda)W_2^{-1} \succcurlyeq (\lambda W_1 + (1-\lambda)W_2)^{-1},
    \end{equation}
    which is equivalent to saying that for any $v\in\mathbb{R}^n$
    
    \begin{equation}
        \lambda v^\top W_1^{-1}v + (1-\lambda)v^\top W_2^{-1}v \geq v^\top(\lambda W_1 + (1-\lambda)W_2)^{-1}v,
    \end{equation}
    
    Defining $\tilde{g}(\lambda) = u^\top(\lambda W_1 + (1-\lambda)W_2)u$ for any positive definite matrices $W_1$ and $W_2$ and nonzero vector $u$ allows us to rewrite the inequality above as.
    
    \begin{equation}
        \lambda \tilde{g}(0) + (1-\lambda)\tilde{g}(1) \geq \tilde{g}(\lambda).
    \end{equation}
    
    Define $Z(\lambda) = \lambda W_1 + (1-\lambda)W_2$ and compute the second derivative of $\tilde{g}$ with respect to $\lambda$ gives
    
    \begin{equation}
        \frac{d^2g}{d\lambda^2} = 2(ZZ^{-1}u)^\top Z^{-1} ZZ^{-1}u = v^\top Z^{-1} v \geq 0
    \end{equation}
    since the inverse of the convex combination of positive definite matrices is positive definite. Therefore, $\tilde{g}(\lambda)$ is convex for any $u$ and any positive definite $W_1$ and $W_2$ for $\lambda \in [0,1]$ which implies $\bar{g}$ is convex and, therefore, $g$ is convex.
    
\end{proof}

With this we can conclude that $f = g\circ h$ is convex since it is the composition of a convex function with an affine one. Since our cost function is convex and our constraints affine, we solve this relaxed problem for a given number $\ell$ of attacked edges using a gradient descent algorithm to get a lower bound on the optimal point as well as an argument vector $c$. By picking the $\ell$ largest directions of $c$ we can obtain a rounded solution as the third method for solving \eqref{eq:optimincardinal}.}






\section{Simulations}
\label{sec:simul}
In this section, we demonstrate our proposed algorithms for different graphs with varying complexities. For each network, we vary our budget from protecting a single edge to protecting all edges, comparing the solutions of each method with the lower bound derived in Subsection \ref{sub:lowbnd} and with the expected gain from randomly selecting the edges to protect. We also assume that every edge of the network is vulnerable, that is $\mathcal{E}_v = \mathcal{E}$.
{
Along this section, the drift matrix $N_0$ in the bilinear dynamics \eqref{eq:BlnNetDyn} will be given by
\begin{equation}
    N_0=-L-\frac{1}{n}\mathds{1}_{n \times n},
    \label{N0:1239}
\end{equation}
where $L$ is the Laplacian matrix of the corresponding graph at each section (c.f. Figs. \ref{fig:ringraph}, \ref{fig:BAGraph} and \ref{fig:AirportNetwork}) and $\mathds{1}$ is the matrix of all ones. For each graphs the nodes indicated in red (as node $1$ in Fig. \ref{fig:ringraph}) are disturbed by an additive attack $v$ (i.e., $B$ is the column composition of elementary vectors $e_i$ for all $i\in\mathcal{V}_a$). Finally, in \eqref{eq:BlnNetDyn}, each $N_k$ corresponds to an unprotected vulnerable edge $(i,j)$ where $(i,j)\in\mathcal{E}_v \backslash \mathcal{E}_p$ and defined by $N_k := E_{ij}+E_{ji}$ where $E_{ij}$ is an elementary matrix as defined in subsection \ref{sub:Notations}.}

Along our simulations, unless specified otherwise, we weight all the $N_k$s by $\bar{\lambda}_2/m$ to make sure Assumption \ref{assump:1} is satisfied, where $m=|\mathcal{E}_v|$ is the number of vulnerable edges, and $\bar{\lambda}_2=\min(1,\lambda_2)$ is the smallest between the smallest nonzero eigenvalue of $L$ and $1$.

\subsection{The Ring Graph}

\begin{figure}[t]
    \centering
    \includegraphics[scale = 1.2]{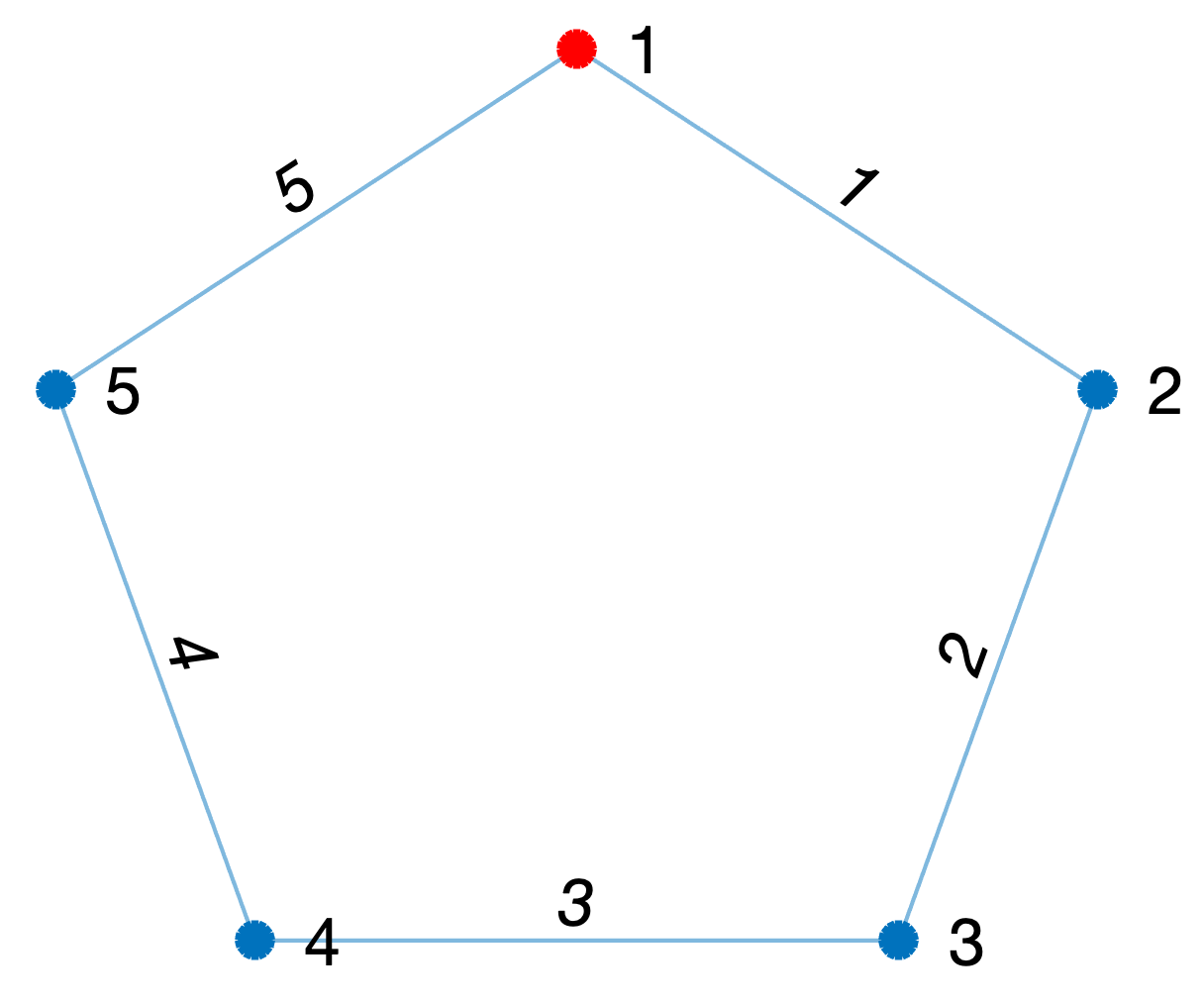}
    \caption{This figure shows the drawing of the undirected five-node ring graph used in the Simulations section. The red node is the one under the effect of additive disturbance, and all edges are vulnerable.}
    \label{fig:ringraph}
\end{figure}



\begin{figure*}[t]
    \centering
    \includegraphics[scale = 0.9]{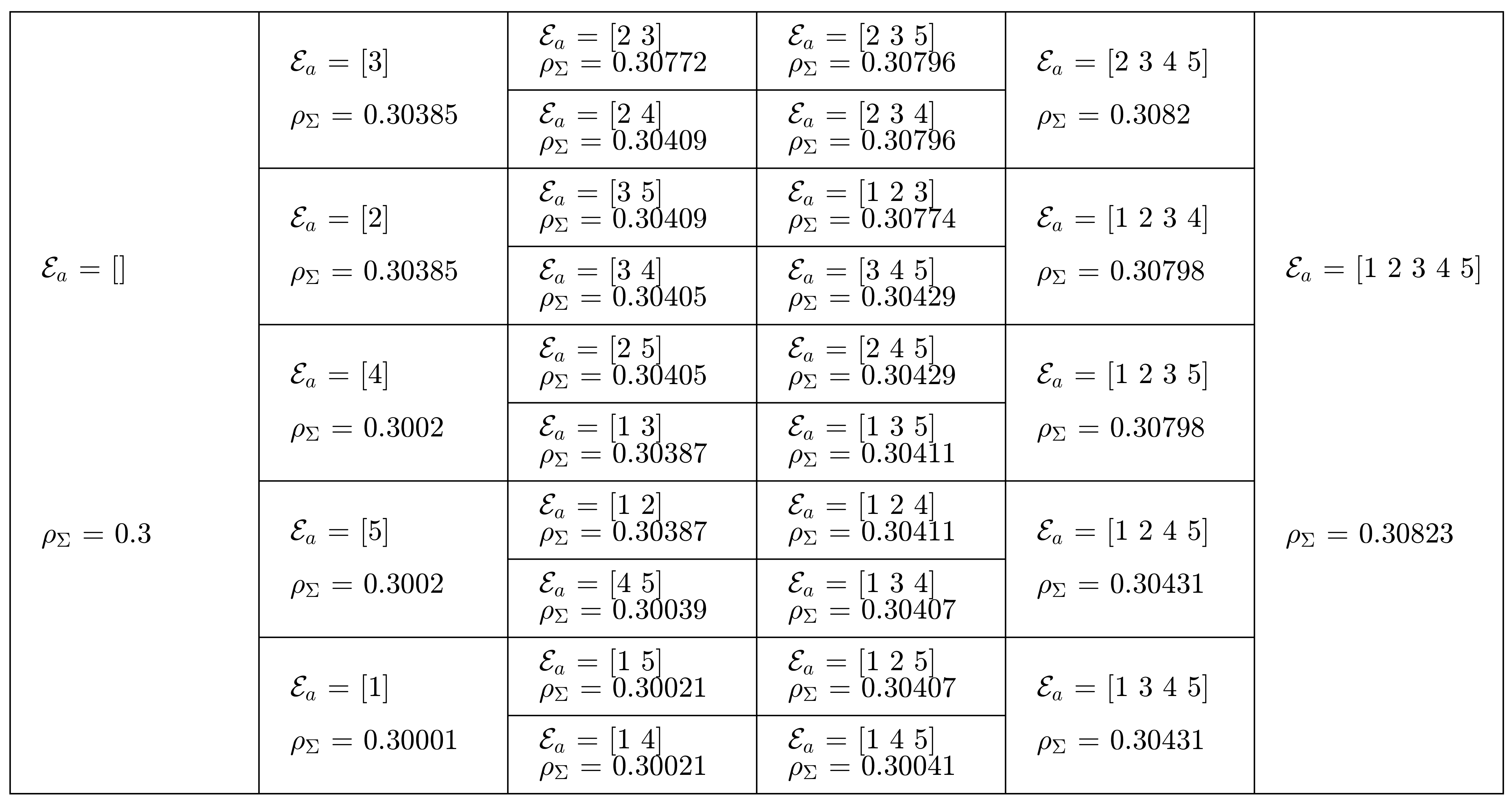}
    \caption{Values of $\rho_\Sigma(\cdot)$ of the 5-node ring graph for every possible subset of attacked edges.}
    \label{fig:H2normgraph}
\end{figure*}

\begin{figure}
    \centering
    \includegraphics[scale=0.7]{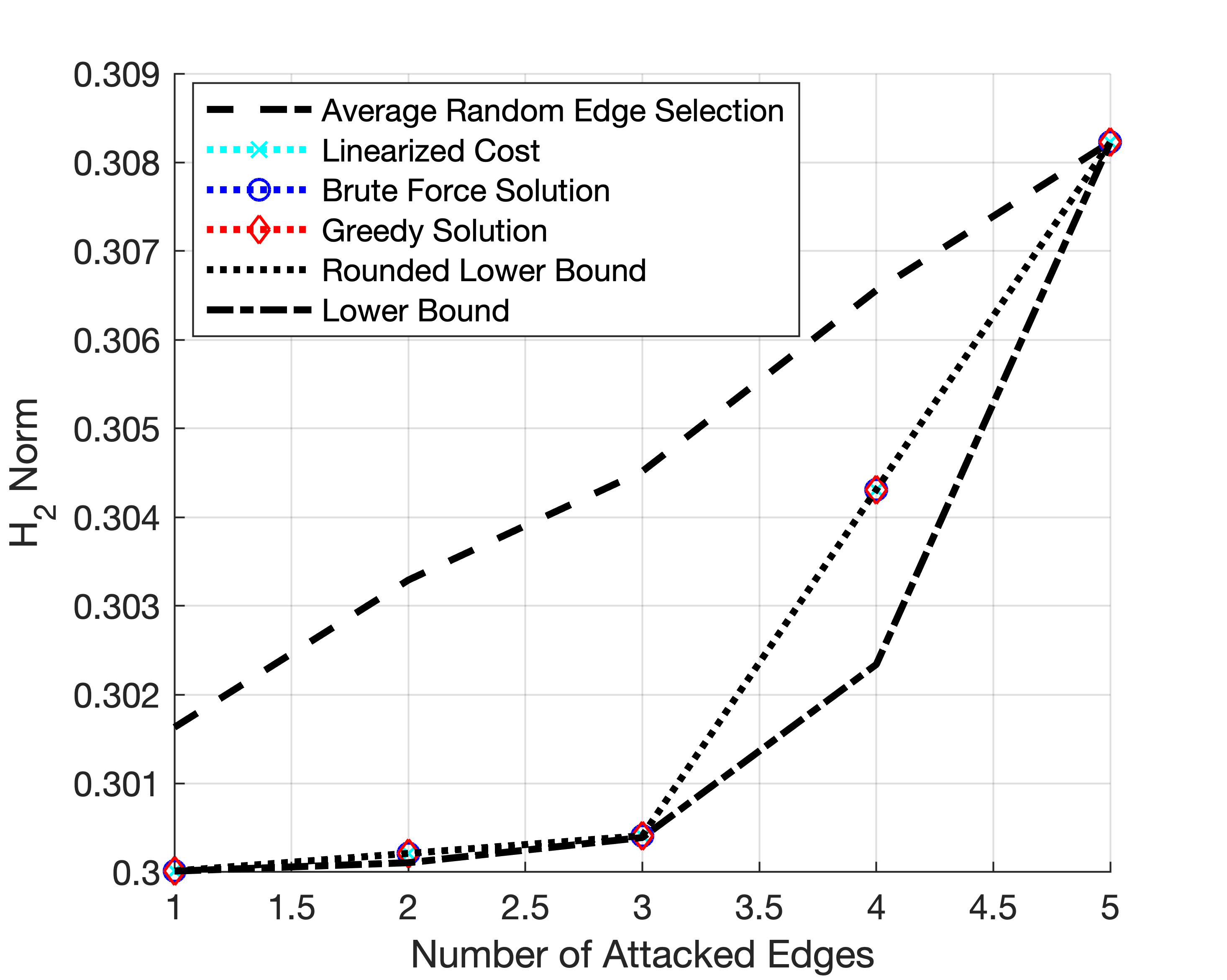}
    \caption{This figure shows our results for a bilinear network with dynamics \eqref{eq:BlnNetDyn} and topology given by a 5-node ring graph. Here we compare the three proposed methods (linearized cost, greedy algorithm and rounding of lower bound solution) for solving the optimization problem \eqref{eq:optimincardinal} with the actual global minimum obtained through brute force.}
    \label{fig:Ring5Results}
\end{figure}

In the first simulation we use a 5-node ring graph labeled as in Fig. \ref{fig:ringraph}. Due to the simplicity of this graph, we can easily compute the $\rho_\Sigma$ function defined in \eqref{eq:rho} for any subset of attacked edges and find the actual optimum through brute force, as shown in Fig.~\ref{fig:H2normgraph}. Notice that, despite the symmetry of the ring graph, some edges have a greater effect on the cost function than others. This happens because of the additive disturbance acting on node one, which introduces an important asymmetry in the system. We can see from Fig. \ref{fig:Ring5Results} that for this system, all proposed algorithms obtained the actual minimum every time.

We can also see that protecting the two most vulnerable edges results in the largest reduction in the value of the $\mathcal{H}_2$-norm. This gives us important insight when planning how to protect a system with this topology cost effectively.

\begin{figure}[h]
    \centering
    \includegraphics[scale=0.66]{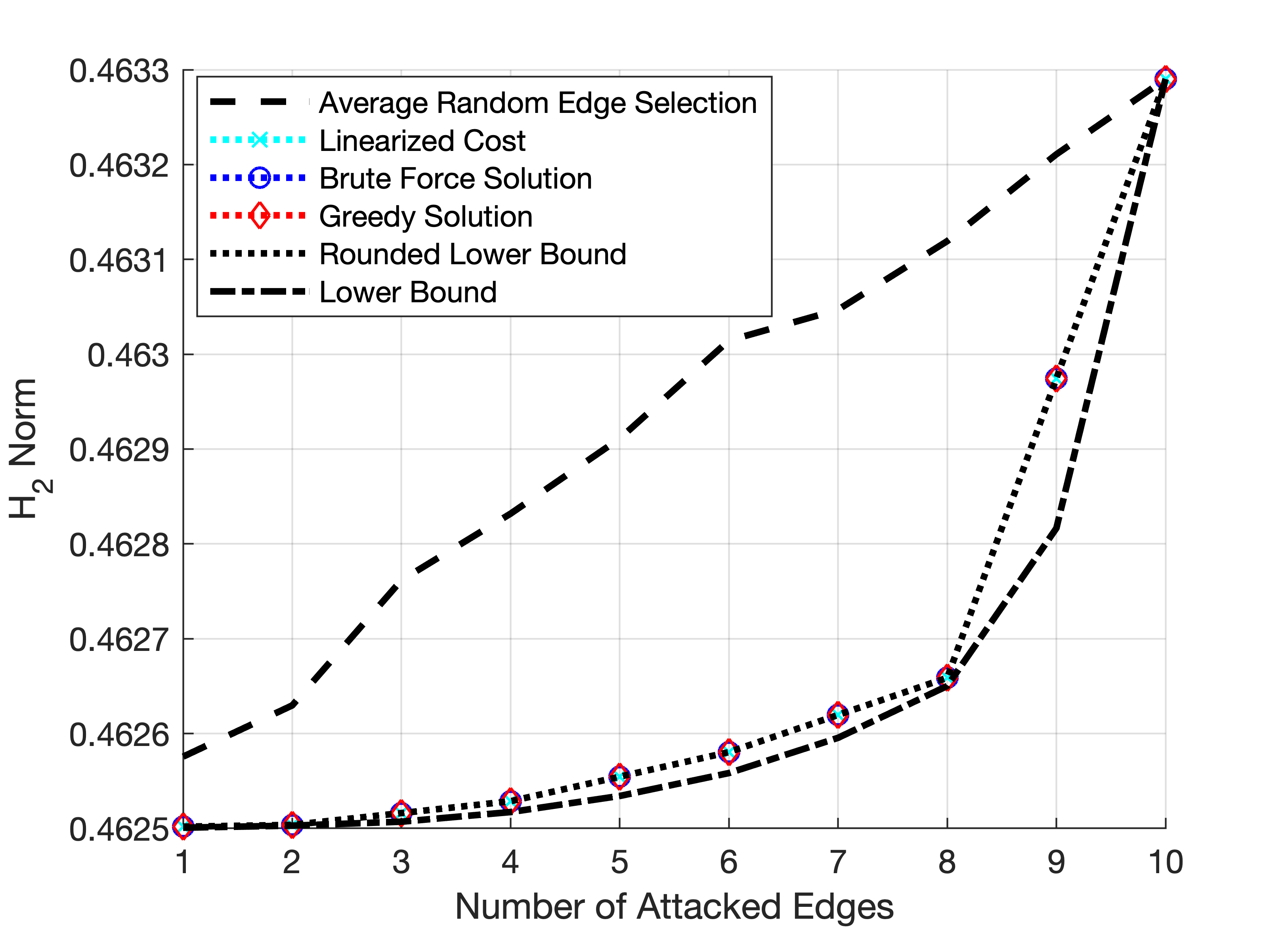}
    \caption{This figure shows our results for a bilinear network with dynamics \eqref{eq:BlnNetDyn} and topology given by a 10-node ring graph. Here we compare the three proposed methods (linearized cost, greedy algorithm and rounding of lower bound solution) for solving the optimization problem \eqref{eq:optimincardinal} with the actual global minimum obtained through brute force.}
    \label{fig:Ring10Results}
\end{figure}

The simulations is repeated for a ring graph with $10$ nodes and $10$ edges, all vulnerable (that is, subejct to multiplicative disturbances if not protected). The results are completely analogous, as can be seen in Fig. \ref{fig:Ring10Results}.

\begin{figure}
    \centering
    \includegraphics[scale=0.66]{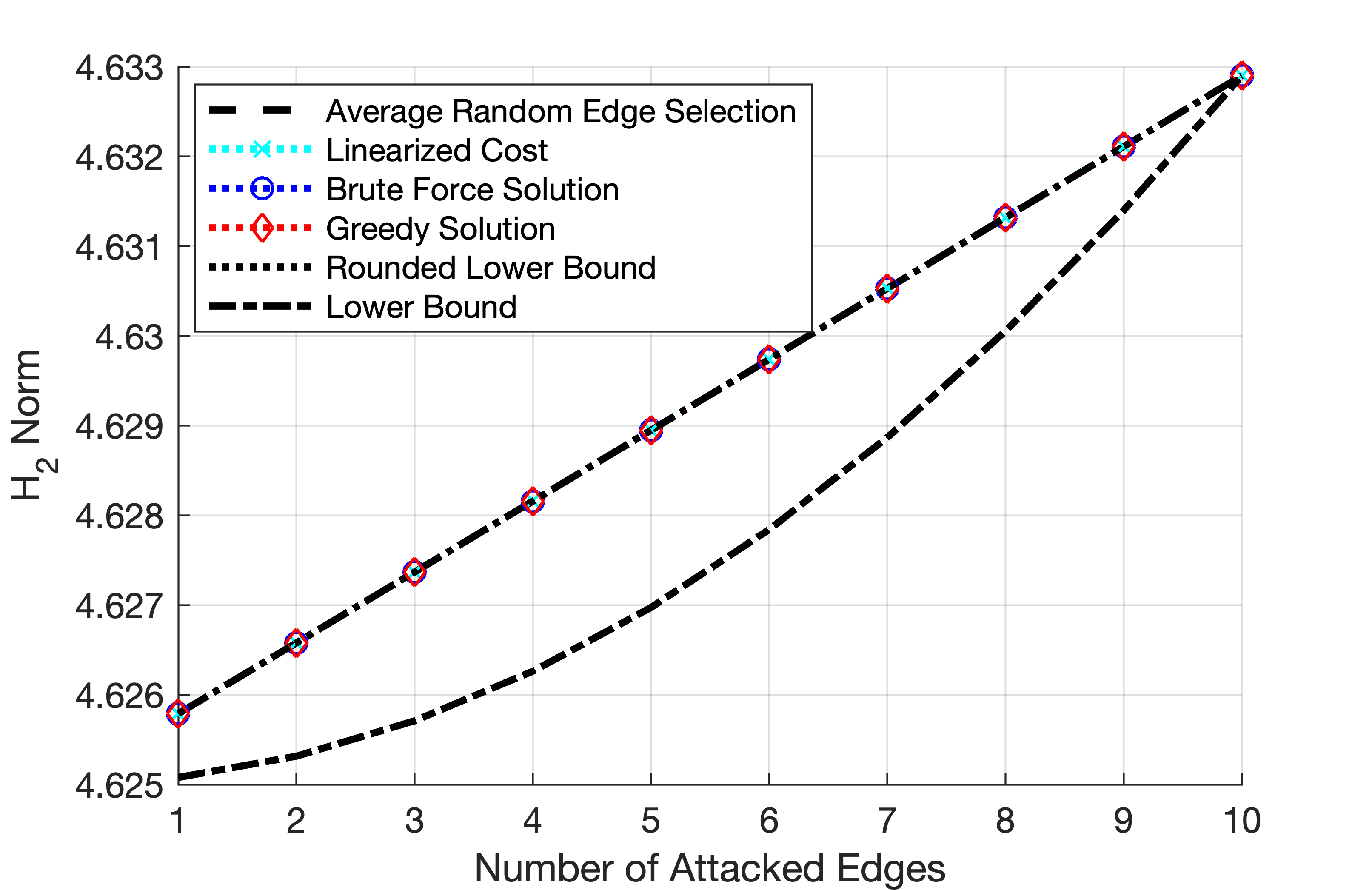}
    \caption{This figure shows our results for a bilinear network with dynamics \eqref{eq:BlnNetDyn} and topology diven by a 10-node ring graph where all nodes are disturbed. In this case we notice that the three proposed methods (linearized cost, greedy algorithm and rounding of lower bound solution) for solving the optimization problem \eqref{eq:optimincardinal} perform exactly as well as the actual global minimum obtained through brute force and as the random edge selection algorithm.}
    \label{fig:Ring10ResultsEye}
\end{figure}

Finally, one last analysis we performed with the ring graph is about the influence of asymmetries of the disturbance distributions on the accuracy of the lower bound approximation. The previous simulations were conducted for a single linear disturbance on Node $1$. In Fig. \ref{fig:Ring10ResultsEye}, we present the result for the simulation of the $10$-node ring graph, but with all nodes under linear disturbances (that is, $B=I_{n\times n}$). Notice that all algorithms (including the random edge selection) performed the same and at the optimum solution. This happened because the perfect symmetry of the graph makes so that all edges have the same influence over the $\mathcal{H}_2$-norm, therefore, which edge to pick at each step does not matter.

\subsection{The Barab\'asi-Albert Random Graph}

\begin{figure}[h]
    \centering
    \includegraphics[scale=1.2]{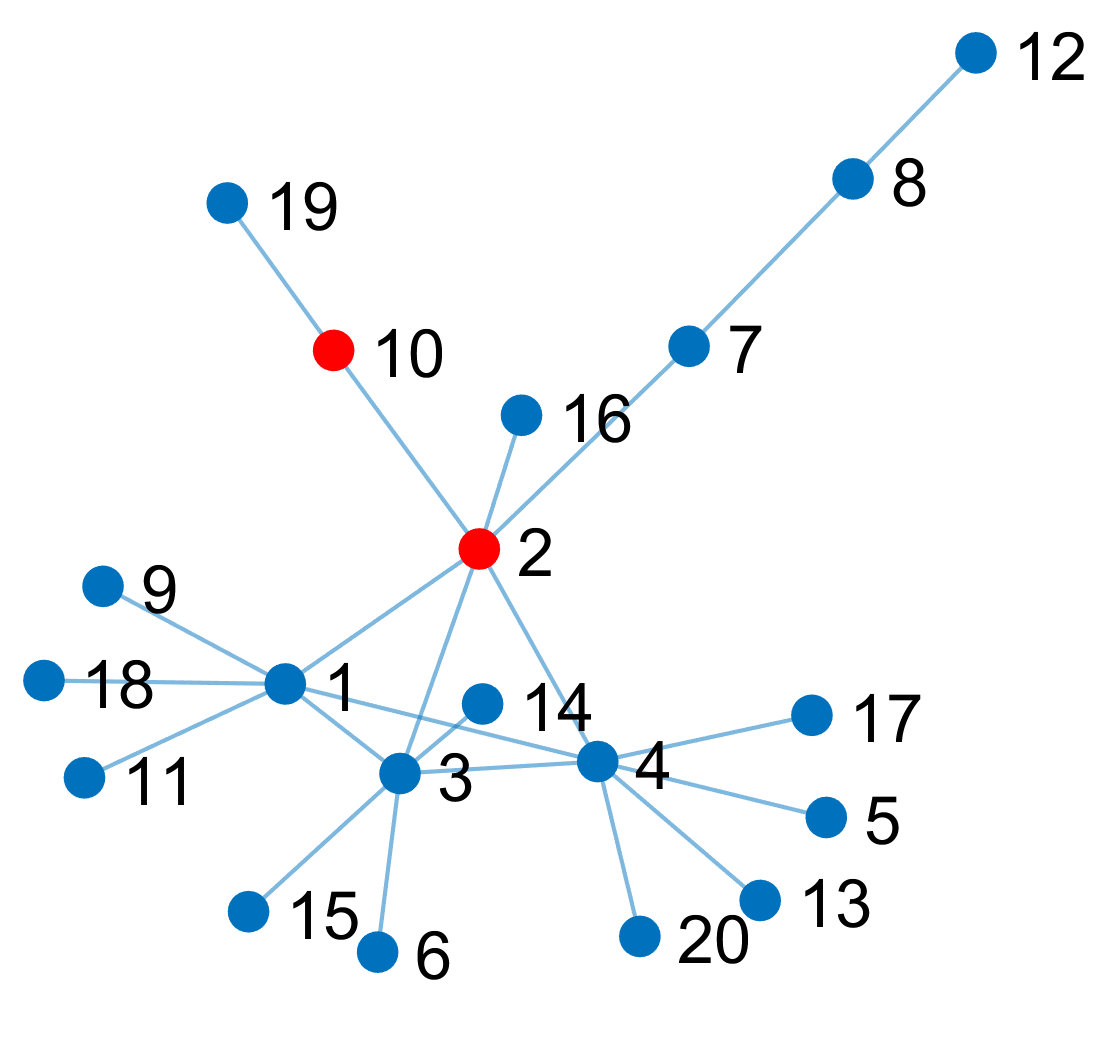}
    \caption{This figure shows the generated Barab\'asi-Albert graph used in our simulations. The red nodes are the one under the effect of additive disturbances, and all edges are vulnerable.}
    \label{fig:BAGraph}
\end{figure}

For a more complex randomly generated network, we simulate the $20$-node Barab\'asi-Albert graph shown in Fig. \ref{fig:BAGraph}. Similarly to the previous simulations, our dynamics are given by \eqref{eq:BlnNetDyn} and the drift matrix by $\eqref{N0:1239}$, where $L$ is the Laplacian for this graph, and the $N_k$'s are given for each unprotected vulnerable edge.

Due to the larger dimension and increasse complexity of the network when compared to the previous simulations ($5$ and $10$ nodes ring graphs), computing the brute force solution becomes prohibitively time-consuming, therefore the lower bound becomes our reference when analysing suboptimality of our results. We also assume, as with the previous simulations, that every edge is independently disturbed in an undirected manner, unless protected. 

\begin{figure}[h]
    \centering
    \includegraphics[scale=0.6]{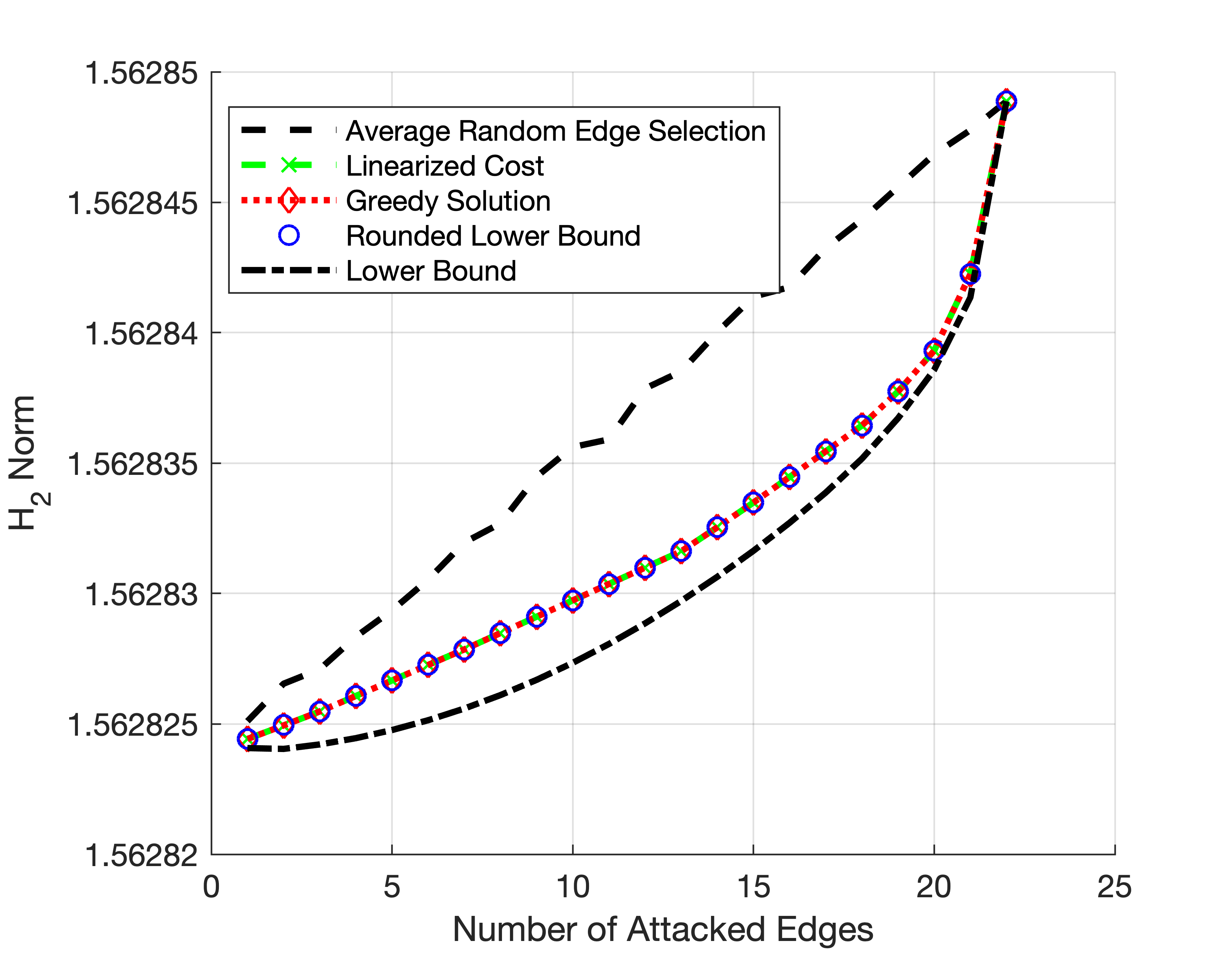}
    \caption{This figure shows our results for a bilinear network with dynamics \eqref{eq:BlnNetDyn} and topology given by the graph shown in Fig. \ref{fig:BAGraph}. Here we compare the three proposed methods (linearized cost, greedy algorithm and rounding of lower bound solution) for solving the optimization problem \eqref{eq:optimincardinal} with our lower bound and the random edge selection. Notice that for this case, where the bilinear matrices $N_k$'s are weighted to respect Assumption \ref{assump:1}, the methods perform identically.}
    \label{fig:BAResults_RA1}
\end{figure}

\begin{figure}[h]
    \centering
    \includegraphics[scale = 0.8]{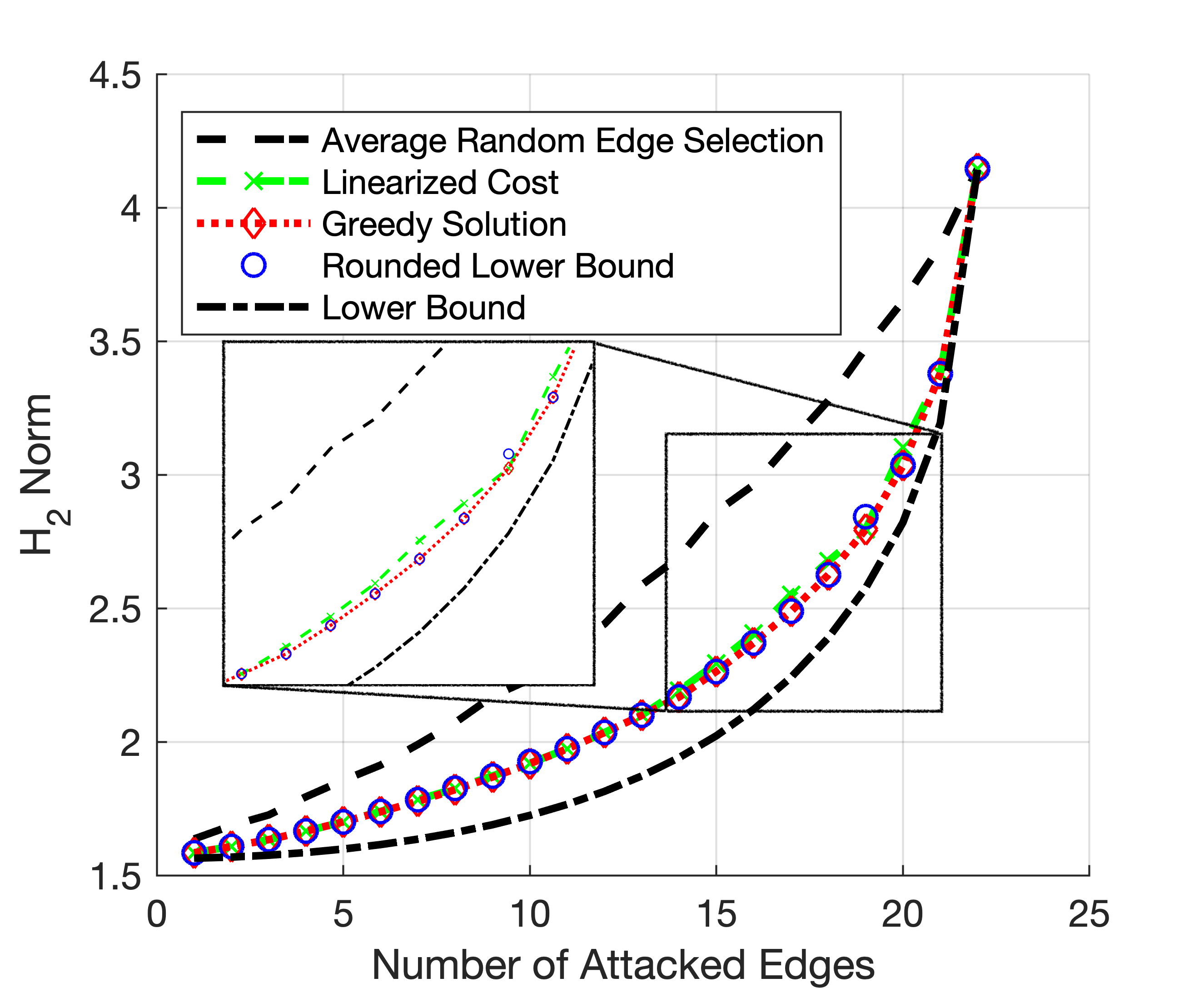}
    \caption{This figure shows our results for a bilinear network with dynamics \eqref{eq:BlnNetDyn} and topology given by the graph shown in Fig. \ref{fig:BAGraph}. Here we compare the three proposed methods (linearized cost, greedy algorithm and rounding of lower bound solution) for solving the optimization problem \eqref{eq:optimincardinal} with our lower bound and the random edge selection. In this simulatons, the bilinear matrices $N_k$'s are unweighted but the dynamics still respects Assumption \ref{assump:2}.}
    \label{fig:BAResults_NRA1}
\end{figure}

We can see from Fig. \ref{fig:BAResults_RA1} that, as it was with the ring graph simulations, the three proposed algorithms have exactly the same results. This is likely due to the fact that Assumption \ref{assump:1} is too restrictive. By asking that the linear dynamics be dominant in the worst case (all vulnerable edges attacked), we make it so that all three algorithms have the same results, as they would if the dynamics were perfectly linear to begin with. To try and observe the effects of the bilinear dynamics on the different algorithms, we run a second simulation where we do not weight the $N_k$'s to respect Assumption \ref{assump:1}, but we verify that it still respects Assumption \ref{assump:2}. The results are presented in Fig. \ref{fig:BAResults_NRA1}.

Notice from the second set of simulations that the solutions based on the linearized cost, rounded lower bound and greedy method differ for a different number of attacked edges, with the greedy edge selection being consistently the best performing algorithm.

\subsection{Airport Network Simulation}

\begin{figure}
    \centering
    \includegraphics[scale=1.2]{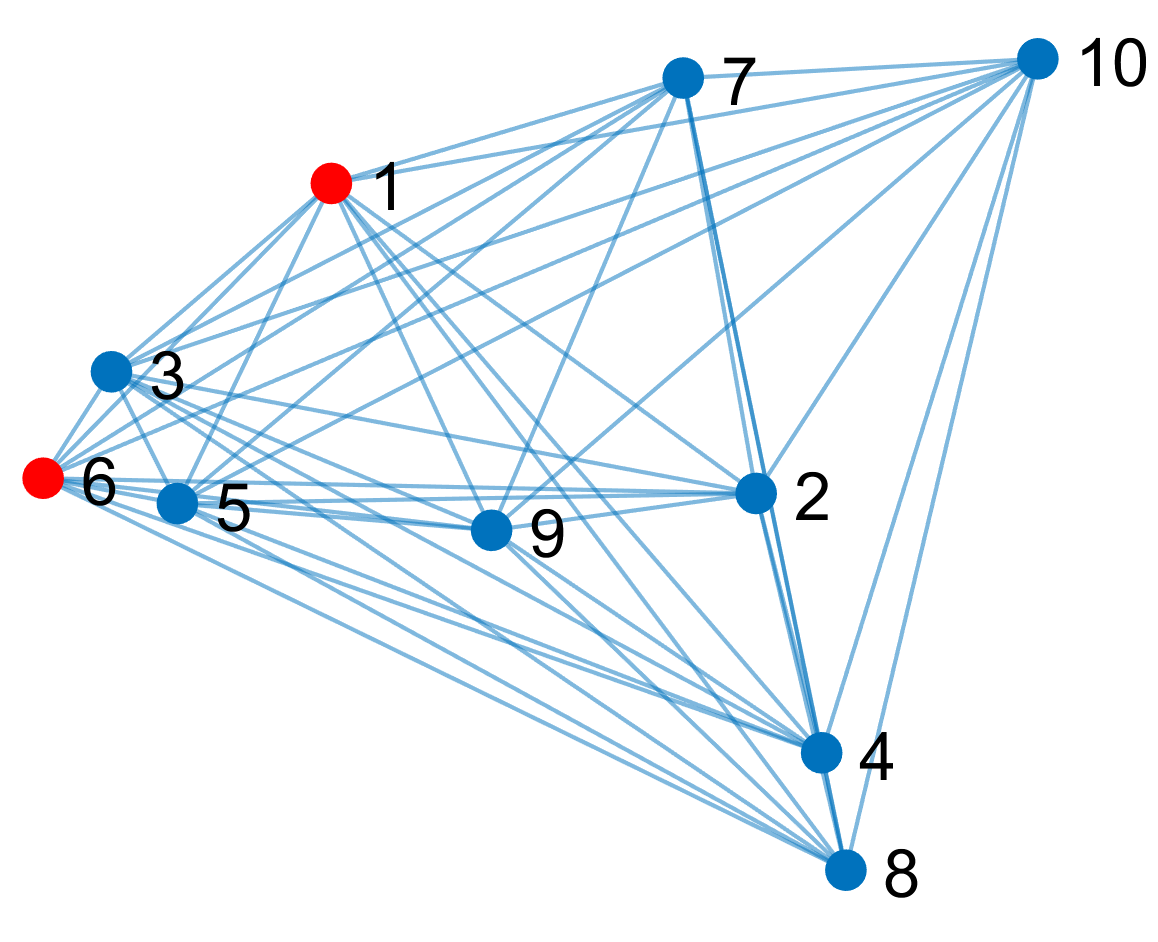}
    \caption{This figure shows the extracted subgraph of the $20$ busiest domestic airports in the USA. Red nodes are the ones under the effect of linear disturbance in the simulations.}
    \label{fig:AirportNetwork}
\end{figure}

The network for our next set of simulations is extracted from the global airports dataset \cite{USdata:Online} by isolating the USA airports and selecting the $20$ with the highest number of connections. As with the previous cases, our dynamics are given by \eqref{eq:BlnNetDyn} and the drift matrix by $\eqref{N0:1239}$. The considered network is presented in Fig. \ref{fig:AirportNetwork}.

Notice that, while the network data is used to evaluate this method has value for being a naturally occurring network topology, we do not claim that an edge disturbance approach makes sense for this specific application.
\begin{figure}[h]
    \centering
    \includegraphics[scale=0.66]{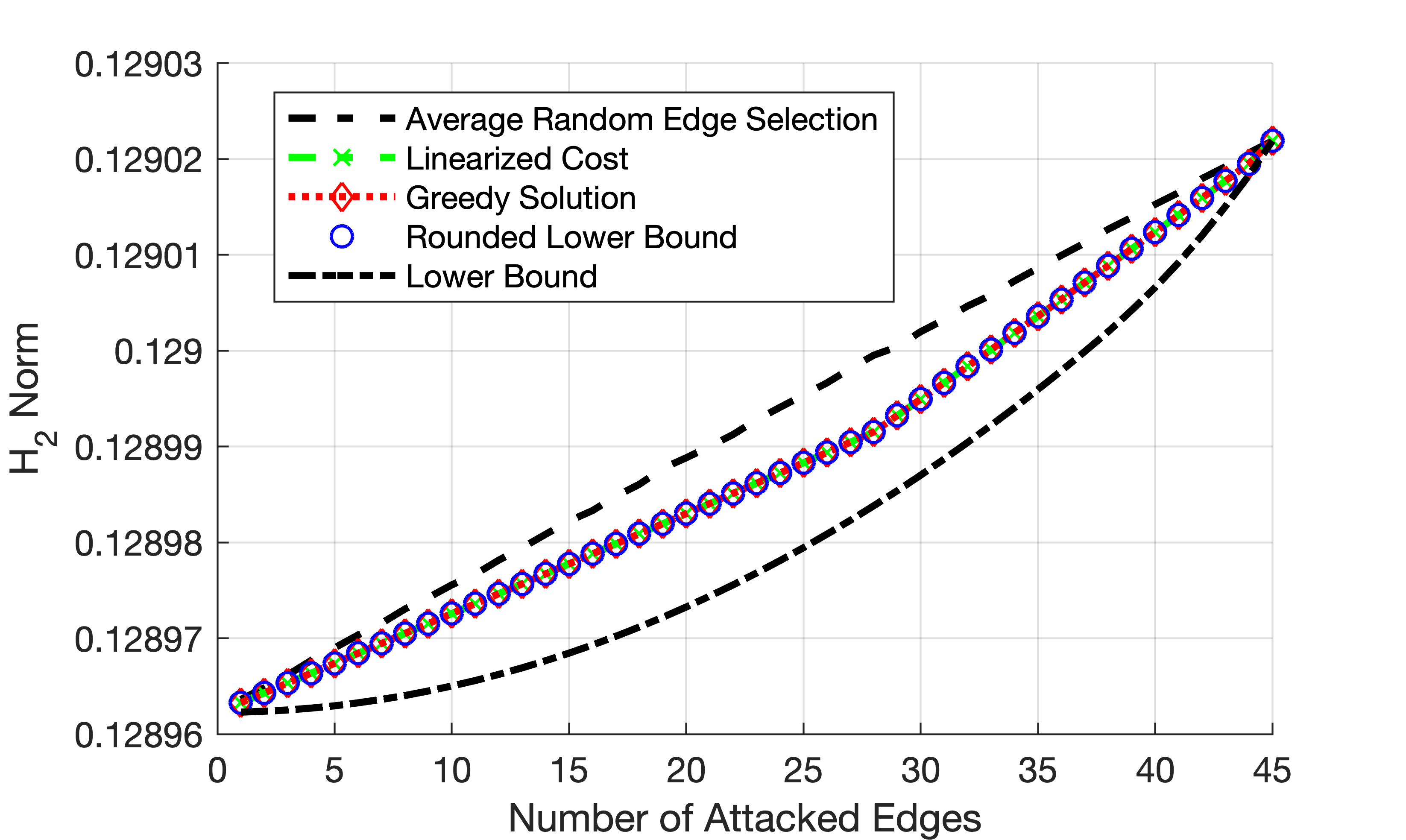}
    \caption{This figure shows our results for a bilinear network with dynamics \eqref{eq:BlnNetDyn} and topology given by the graph shown in Fig. \ref{fig:AirportNetwork}. Here we compare the three proposed methods (linearized cost, greedy algorithm and rounding of lower bound solution) for solving the optimization problem \eqref{eq:optimincardinal} with our lower bound and the random edge selection. Notice that for this case, where the bilinear matrices $N_k$'s are weighted to respect Assumption \ref{assump:1}, the methods perform identically.}
    \label{fig:SimulAirport}
\end{figure}

\begin{figure}[h]
    \centering
    \includegraphics[scale=0.66]{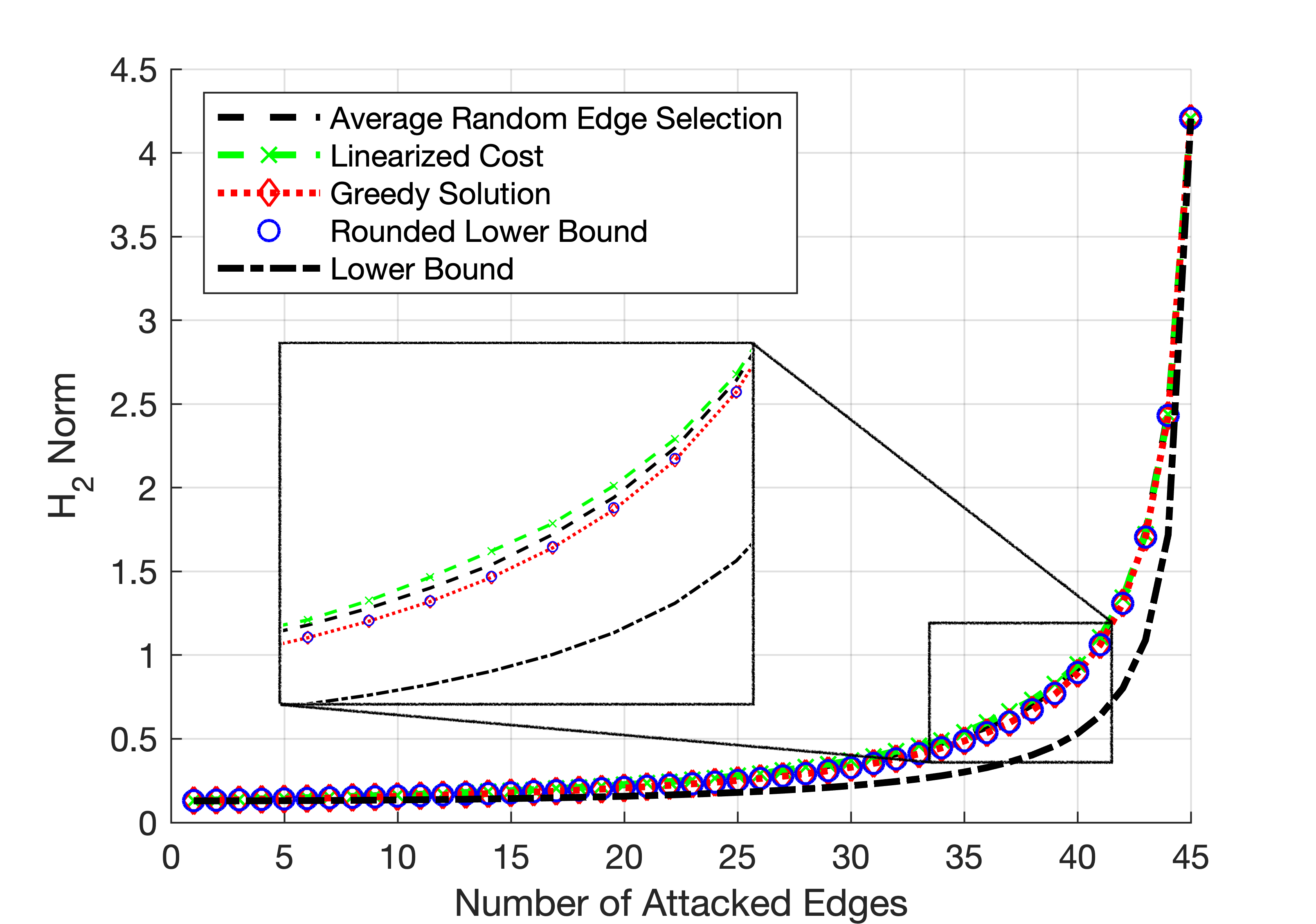}
    \caption{This figure shows our results for a bilinear network with dynamics \eqref{eq:BlnNetDyn} and topology given by the graph shown in Fig. \ref{fig:AirportNetwork}. Here we compare the three proposed methods (linearized cost, greedy algorithm and rounding of lower bound solution) for solving the optimization problem \eqref{eq:optimincardinal} with our lower bound and the random edge selection. In this simulatons, the bilinear matrices $N_k$'s are unweighted but the dynamics still respects Assumption \ref{assump:2}.}
    \label{fig:SimulAirportNRA1}
\end{figure}

As we can see in Fig. \ref{fig:SimulAirport}, while all algorithms are better than a random edge selection, they, once again, all perform the same for the case where Assumption \ref{assump:1} is respected. We next investigate a case where the $N_k$'s are unweighted but Assumption \ref{assump:2} is still satisfied. The resulting simulation is presented in Fig. \ref{fig:SimulAirportNRA1}.

As we can see, similarly to the Barab\'asi-Albert graph, once we relax Assumption \ref{assump:1} to \ref{assump:2}, the discrepancies on the algorithms become more relevant. Namely, the method based on the linearized cost starts performing worse than even the random edge selection. The second point of interest is that the greedy and round lower bound solutions, although strictly better than the random edge selection, obtain results numerically close to it. We hypothesise that this effect is due to the density of the considered graph, making so that the edges have similar weights.

\section{Conclusions and Future Works}
\label{sec:conclusions}

In this paper, we proposed a way to evaluate the influence of multiplicative disturbances to the overall stability of the system by making use of the $\mathcal{H}_2$-norm defined for bilinear systems. We discussed the meanings of the $\mathcal{H}_2$-norm and why it is an interesthing metric, and showed that in the context of edge selection, it is supermodular, which allows us to use efficient selection algorithms with guaranteed known optimality gaps. Important to our $\mathcal{H}_2$-based edge selection method, we discussed how Assumption \ref{assump:1} might be too restrictive by, in practice, requiring the dominance of the linear dynamics over the bilinear one. Furthermore, we discuss a possible relaxation of that assumption, originally derived for Gaussian disturbances, where the $\mathcal{H}_2$-norm still maintains its relationship to the steady-state covariance matrix of the states. 

On a more practical note, we proposed a general lower bound for the optimal solution of the combinatorial edge selection problem through its continuous relaxation. We simulated three network topologies: one of a simple ring graph, chosen so that we could compare both the greedy solution and the lower bound to the actual minimum; one of a more complex $20$ node Barab\'asi-Albert graph; and one of the $20$ busiest domestic airports in the USA. In the simulations we compare the greedy algorithm solution with two others: one obtained from linearizing the cost function; and one obtained from rounding the lower bound solution. While the three algorithms performed really similarly, the greedy solution was the lowest one for all the simulated cases.

We believe that our edge-perturbation results will be of interest in other areas besides classical optimization ones such as transportation networks.  In fact, we were largely motivated to pursue this work by the analysis of the impact of edge knock-outs in biological networks carried out in~\cite{bleris2018edges}, which used CRISPR technology to target microRNA pathways that control processes ranging from cell growth to stress responses.  That paper emphasized how edge perturbations play a critical role in cell function, analogously to how node perturbations lead to global behavioral changes through network effects~\cite{15bleris_direct_indirect}.




\bibliographystyle{IEEEtran}
\bibliography{EdgeSelectionBibliography.bib}{}

\section*{Appendix}
    \subsection*{Proof of Theorem \ref{thm:mainresult}}
        To prove Theorem \ref{thm:mainresult}, we first prove the following three Lemmas.
        
        \begin{lemma}
            \label{lm:pqfactorization}
            Let us define $\tilde{P}_{q}(\mathcal{E}_a, \bar{t}_q) = P_q(\mathcal{E}_a, \bar{t}_q)P_q^\top(\mathcal{E}_a, \bar{t}_q)$, where $\bar{t}_q = [t_1, \dots, ~t_q]$ and $P_qP_q^\top$ is given by \eqref{eq:PqPqTGen}. Then for $q\geq 2$, $\mathcal{A}$, $\mathcal{B}\subset\mathcal{E}_v$, $\mathcal{A}\cap\mathcal{B}=\emptyset$, we have
            
            \begin{equation}
                \small
                \tilde{P}_q(\mathcal{A}\cup \mathcal{B}, \bar{t}_q) = \tilde{P}_q(\mathcal{A}, \bar{t}_q) + \tilde{P}_q(\mathcal{B}, \bar{t}_q) + \tilde{R}_q(\mathcal{A}, \mathcal{B}, \bar{t}_q),
            \end{equation}
            where $\tilde{R}_q(\mathcal{A}, \mathcal{B}, \bar{t}_q)\succcurlyeq 0$
        \end{lemma}
        
        \begin{proof}
            We prove the Lemma by induction. First notice that for $q=1$
            
            \begin{equation}
                \small
                \tilde{P}_1(\mathcal{A}\cup\mathcal{B}, \bar{t}_1) = e^{N_0t_1}BB^\top e^{N_0^\top t_1},
            \end{equation}
            
            \noindent is constant and independent from the set of attacked edges, so the relationship from the lemma does not hold for $q=1$ since $\tilde{P}_1(\mathcal{A}\cup\mathcal{B}, \bar{t}_1) = \tilde{P}_1(\mathcal{A}, \bar{t}_1) = \tilde{P}_1(\mathcal{B}, \bar{t}_1) = \tilde{P}_1(\bar{t}_1)$. For $q\geq 2$ we can write
            
            \begin{equation}
            \small
            \begin{split}
                \tilde{P}_2(\mathcal{A}\cup\mathcal{B}, \bar{t}_2) &= e^{N_0t_2} \sum_{k\in\mathcal{A}\cup\mathcal{B}}N_k\tilde{P}_1(\bar{t}_1)N_k^\top e^{N_0^\top t_2} = \\ &=e^{N_0t_2} \sum_{k\in\mathcal{A}}N_k\tilde{P}_1(\bar{t}_1)N_k^\top e^{N_0^\top t_2}+\\&+e^{N_0t_2} \sum_{k\in\mathcal{B}}N_k\tilde{P}_1(\bar{t}_1)N_k^\top e^{N_0^\top t_2} = \\&= \tilde{P}_2(\mathcal{A}, \bar{t}_2)+\tilde{P}_2(\mathcal{B}, \bar{t}_2)
            \end{split}
            \end{equation}
            
            \noindent which proves the base case, since $\tilde{R}_2=0$ is positive semidefinite. Furthermore, if we assume the following
            
            \begin{equation}
            \small
                    \begin{split} \tilde{P}_{q-1}(\mathcal{A}\cup\mathcal{B}, \bar{t}_{q-1}) &= \tilde{P}_{q-1}(\mathcal{A}, \bar{t}_{q-1})+\\&+\tilde{P}_{q-1}(\mathcal{B}, \bar{t}_{q-1})+\\&+\tilde{R}_{q-1}(\mathcal{A}, \mathcal{B}, \bar{t}_{q-1})
                \end{split}
            \end{equation}
            with $\tilde{R}_{q-1}\succcurlyeq 0$, then
            \begin{equation}
            \small
            \begin{split}
                \tilde{P}_q(\mathcal{A}\cup\mathcal{B}, \bar{t}_q) &= e^{N_0t_q}\Bigg(\sum_{k\in\mathcal{A}}N_k\tilde{P}_{q-1}(\mathcal{A}\cup\mathcal{B}, \bar{t}_{q-1})N_k^\top+\\&+ \sum_{k\in\mathcal{B}}N_k\tilde{P}_{q-1}(\mathcal{A}\cup\mathcal{B}, \bar{t}_{q-1})N_k^\top \Bigg)e^{N_0^\top t_q} =\\
                &= \tilde{P}_q(\mathcal{A}, \bar{t}_q)+\tilde{P}_q(\mathcal{B}, \bar{t}_q)+\\&+e^{N_0t_q}\Bigg(\sum_{k\in\mathcal{A}}N_k\tilde{P}_{q-1}(\mathcal{B}, \bar{t}_{q-1})N_k^\top+\\&+\sum_{k\in\mathcal{B}}N_k\tilde{P}_{q-1}(\mathcal{A}, \bar{t}_{q-1})N_k^\top+\\&+\sum_{k\in\mathcal{A}\cup\mathcal{B}}N_k\tilde{R}_{q-1}(\mathcal{A},\mathcal{B}, \bar{t}_{q-1})N_k^\top \Bigg)e^{N_0^\top t_q}=\\&=\tilde{P}_q(\mathcal{A}, \bar{t}_q)+\tilde{P}_q(\mathcal{B}, \bar{t}_q)+R_{q}(\mathcal{A},\mathcal{B}, \bar{t}_q)
            \end{split}
            \end{equation}
            
            To finish the induction step we can verify that $R_{q}$ is positive semi-definite since it is assumed $R_{q-1}\succcurlyeq 0$, since positive semi-definiteness is invariant under congruence transformations and matrix addition, and since $\tilde{P}_q\succcurlyeq 0$, then $R_q\succcurlyeq 0$. This completes the proof.
        \end{proof}
        
        \begin{lemma}
            \label{lm:Pfactorization}
            Given $\mathcal{A}$, $\mathcal{B}\subset\mathcal{E}_v$, $\mathcal{A}\cap\mathcal{B}=\emptyset$, and $P$ defined by \eqref{eq:gramiandef}, the associated Gramian $P(\mathcal{A}\cup \mathcal{B})$ can be rewritten as follows
            
            \begin{equation}
            \small
                P(\mathcal{A}\cup \mathcal{B}) = P(\mathcal{A})+P(\mathcal{B})+R(\mathcal{A}, \mathcal{B}) -C
            \end{equation}
            where $R(\mathcal{A}, \mathcal{B})\succcurlyeq 0$, and $C = \int_{0}^\infty e^{N_0t}BB^\top e^{N_0^\top t}dt\succcurlyeq 0$ is constant matrix that is independent of the set of attacked edges.
        \end{lemma}
        \begin{proof}
            Applying Lemma \ref{lm:pqfactorization} to \eqref{eq:gramiandef}, we get
            \begin{equation}
                \small
                \label{eq:Rdef}
                \begin{split}
                    P(\mathcal{A}\cup\mathcal{B}) &= \sum_{q=1}^\infty\int_0^\infty\dots\int_0^\infty\tilde{P}_q(\mathcal{A}\cup\mathcal{B}, \bar{t}_q)dt_1\dots dt_q =\\
                    & = \sum_{q=2}^\infty\int_0^\infty\dots\int_0^\infty\Bigg(\tilde{P}_q(\mathcal{A}, \bar{t}_q)+\tilde{P}_q(\mathcal{B}, \bar{t}_q)+\\&~~~~~~~~~~~~~~~~~~~~+\tilde{R}_q(\mathcal{A},\mathcal{B}, \bar{t}_q)\Bigg)dt_1\dots dt_q + \\ &~~~~~~+ 2\int_{0}^\infty \tilde{P}_1(\bar{t}_1)dt_1  - \int_{0}^\infty \tilde{P}_1(\bar{t}_1)dt_1 \\&= P(\mathcal{A})+P(\mathcal{B})+\\&~~~+\underbrace{\sum_{q=2}^\infty\int_0^\infty\dots\int_0^\infty\tilde{R}_q(\mathcal{A},\mathcal{B}, \bar{t}_q)dt_1\dots dt_q}_{R(\mathcal{A},\mathcal{B})}-\\&~~~- \int_{0}^\infty \tilde{P}_1(\bar{t}_1)dt_1.
                \end{split}
            \end{equation}
            
            The positive semi-definiteness of $R$ is immediate from the fact that such property is invariant under matrix addition.
        \end{proof}
        
        From this point forward, the dependency of $\tilde{P}_q$ on $\bar{t}_q$ and similar functions is suppressed for compactness of the equations.
        
        \begin{lemma}
            \label{lm:Rmonotonic}
            The function $R$ given by Lemma \ref{lm:Pfactorization} is monotonic. That is, given $\mathcal{A}$, $\mathcal{B}$, $\mathcal{C}\subset\mathcal{E}_v$, all disjoint sets, then
            \begin{equation}
                R(\mathcal{A}\cup \mathcal{C}, \mathcal{B}) \succcurlyeq R(\mathcal{A}, \mathcal{B}).
            \end{equation}
        \end{lemma}
        
        \begin{proof}
            The Lemma is true if the following holds
            \begin{equation}
            \small
                \label{eq:Rfactorization}
                R(\mathcal{A}\cup\mathcal{C}, \mathcal{B}) = R(\mathcal{A}, \mathcal{B}) + U(\mathcal{A}, \mathcal{B}, \mathcal{C})
            \end{equation}
            \noindent with $U\succcurlyeq0$, since if it holds we can rewrite the inequality for monotonicity as 
            \begin{equation}
            \small
                 R(\mathcal{A}, \mathcal{B}) + U(\mathcal{A}, \mathcal{B}, \mathcal{C})-R(\mathcal{A}, \mathcal{B}) = U(\mathcal{A}, \mathcal{B}, \mathcal{C})\succcurlyeq 0.
            \end{equation}
            To show that \eqref{eq:Rfactorization} holds it is enough to show that 
            \begin{equation}
            \small
                \tilde{R}_q(\mathcal{A}\cup\mathcal{C}, \mathcal{B}) = \tilde{R}_q(\mathcal{A}, \mathcal{B}) + \tilde{U}_q(\mathcal{A}, \mathcal{B}, \mathcal{C})
            \end{equation}
            \noindent with $\tilde{U}_q\succcurlyeq 0$ holds for all $\tilde{R}_q$s that compose $R$ in \eqref{eq:Rdef}. For $q=2$, $\tilde{R}_2=0$, which holds the inequality trivially, proving the base case of indution. For an arbitrary $q$ assuming it holds for $q-1$, we have
            \begin{equation}
            \small
                \begin{split}
                    \tilde{R}_q(\mathcal{A}&\cup\mathcal{C},\mathcal{B}) = \\=&~ e^{N_0t_q}\Bigg(\sum_{k\in\mathcal{A}\cup\mathcal{C}}N_k\tilde{P}_{q-1}(\mathcal{B})N_k^\top+\\&~~~~~~~~+\sum_{k\in\mathcal{B}}N_k\tilde{P}_{q-1}(\mathcal{A}\cup\mathcal{C})N_k^\top+\\&+~~~~\sum_{k\in\mathcal{A}\cup\mathcal{C}\cup\mathcal{B}}N_k\tilde{R}_{q-1}(\mathcal{A}\cup\mathcal{C},\mathcal{B})N_k^\top \Bigg)e^{N_0^\top t_q} \\
                    =&~ e^{N_0t_q}\Bigg(\sum_{k\in\mathcal{A}}N_k\tilde{P}_{q-1}(\mathcal{B})N_k^\top+\\&~~~~~~~~+\sum_{k\in\mathcal{C}}N_k\tilde{P}_{q-1}(\mathcal{B})N_k^\top+\\&~~~~~~~~+\sum_{k\in\mathcal{B}}N_k\bigg(\tilde{P}_{q-1}(\mathcal{A})+\\&~~~~~~~~~~~~~~~~~~+\tilde{P}_{q-1}(\mathcal{C})+\\&~~~~~~~~~~~~~~~~~~+\tilde{R}_{q-1}(\mathcal{A}, \mathcal{C})\bigg)N_k^\top+\\&~~~~~~~+\sum_{k\in\mathcal{A}\cup\mathcal{B}}N_k\bigg(\tilde{R}_{q-1}(\mathcal{A},\mathcal{B})+\\&~~~~~~~~~~~~~~~~~~+\tilde{U}_{q-1}(\mathcal{A}, \mathcal{B}, \mathcal{C})\bigg)N_k^\top +\\&~~~~~~~~~+\sum_{k\in\mathcal{C}}N_k\tilde{R}_{q-1}(\mathcal{A}\cup\mathcal{C},\mathcal{B})N_k^\top\Bigg)e^{N_0^\top t_q} \\ =&~ \tilde{R}_q(\mathcal{A}, \mathcal{B})+e^{N_0t_q}\Bigg(\sum_{k\in\mathcal{C}}N_k\tilde{P}_{q-1}(\mathcal{B})N_k^\top+\\&+\sum_{k\in\mathcal{B}}N_k\left(\tilde{P}_{q-1}(\mathcal{C})+\tilde{R}_{q-1}(\mathcal{A}, \mathcal{C})\right)N_k^\top+\\&+\sum_{k\in\mathcal{A}\cup\mathcal{B}}N_k\left(\tilde{U}_{q-1}(\mathcal{A}, \mathcal{B}, \mathcal{C})\right)N_k^\top+\\&+\sum_{k\in\mathcal{C}}N_k\tilde{R}_{q-1}(\mathcal{A}\cup\mathcal{C},\mathcal{B}, \bar{t}_{q-1})N_k^\top \Bigg)e^{N_0^\top t_q} = \\ &= \tilde{R}_q(\mathcal{A}, \mathcal{B}) + \tilde{U}_q(\mathcal{A}, \mathcal{B}, \mathcal{C}),
                \end{split}
            \end{equation}
            \noindent completing the induction step of the proof (the positive semidefiniteness of $\tilde{U}_q$ is done in exactly the same way as for $\tilde{R}_q$ in Lemma \ref{lm:Pfactorization}).
            
        \end{proof}
        
        Using Lemmas \ref{lm:pqfactorization}, \ref{lm:Pfactorization}, and \ref{lm:Rmonotonic}, we prove Theorem \ref{thm:mainresult} as follows.
        
        \begin{proof}
            A set function $\rho_\Sigma:2^{\mathcal{E}_v}\rightarrow \mathbb{R}^{+}$ is supermodular if and only if it satisfies
            
            \begin{equation}
            \label{eq:sprmodproof}
            \small
                \label{eq:sprmodproof}
                \rho_{\Sigma}(\mathcal{B}\cup \{e\})-\rho_{\Sigma}(\mathcal{B}) \geq \rho_{\Sigma}(\mathcal{A}\cup \{e\})-\rho_{\Sigma}(\mathcal{A}).
            \end{equation}
            for every $\mathcal{A}$, $\mathcal{B}\subset \mathcal{E}_v$ with $\mathcal{A}\subset\mathcal{B}$ and every $e\in\mathcal{E}_v/\mathcal{B}$. From Lemma \ref{lm:Pfactorization}, we can rewrite the left hand side of inequality \eqref{eq:sprmodproof} as
            \begin{equation}
            \small
                \rho_{\Sigma}(\mathcal{B}\cup \{e\})-\rho_{\Sigma}(\mathcal{B}) = \rho_\Sigma(e)+\trace(R(\mathcal{B}\cup e)),
            \end{equation}
            and the right hand side similarly. Then, (\ref{eq:sprmodproof}) can be simplified as
            \begin{equation}
            \label{eq:rineq}
            \small
                \trace(R(\mathcal{B}\cup e))\geq \trace(R(\mathcal{A}\cup e)),
            \end{equation}
            where $R$ is defined as in Lemma \ref{lm:Rmonotonic}. Inequality \eqref{eq:rineq} always holds for our function because $\mathcal{A}\subset\mathcal{B}$ and Lemma \ref{lm:Rmonotonic}.
            
        \end{proof}

 \begin{IEEEbiography}[{\includegraphics[width=1in,height=1.25in,clip,keepaspectratio, trim= 100 100 100 100 ]{?}}]{Arthur Castello B. de Oliveira} received his B.Sc. in electrical engineering and M.Sc. in control engineering degrees from the University of São Paulo in 2016 and 2018, respectively. He is currently a PhD student at Northeastern University working on analysis and control of bilinear Networks.

His research interests include Nonlinear Systems, Robust and Optimal Control, Robotics and Mechatronic Systems, and Control of Networks.
 \end{IEEEbiography}

 \begin{IEEEbiography}[{\includegraphics[width=1in,height=1.25in,clip,keepaspectratio, trim= 0 0 0 0 ]{Siami}}]{Milad Siami} 
 (S'12-M'18) received his dual B.Sc. degrees in electrical engineering and pure mathematics from Sharif University of Technology in 2009, M.Sc. degree in electrical engineering from Sharif University of Technology in 2011. He received his M.Sc. and Ph.D. degrees in mechanical engineering from Lehigh University in 2014 and 2017 respectively. He was a postdoctoral associate in the Institute for Data, Systems, and Society at MIT, from 2017 to 2019. He is currently an Assistant Professor with the Department of Electrical \& Computer Engineering, Northeastern University, Boston, MA.
 His research interests include distributed control systems, distributed optimization, and applications of fractional calculus in engineering. 
 \end{IEEEbiography}
 
  \begin{IEEEbiography}[{\includegraphics[width=1in,height=1.25in,clip,keepaspectratio, trim= 100 100 100 100 ]{?}}]{Eduardo D. Sontag} received his Licenciado in Mathematics at the University of Buenos Aires (1972) and a Ph.D. in Mathematics (1977) under Rudolf E. Kalman at the University of Florida. From 1977 to 2017, he was at Rutgers University, where he was a Distinguished Professor of Mathematics and a Member of the Graduate Faculty of the Departments of Computer Science of Electrical and Computer Engineering and the Cancer Institute of NJ. He directed the undergraduate Biomathematics Interdisciplinary Major, the Center for Quantitative Biology, and was Graduate Director at the Institute for Quantitative Biomedicine. In January 2018, Dr. Sontag became a University Distinguished Professor in the Departments of Electrical and Computer Engineering and of BioEngineering at Northeastern University, where he is also affiliated with the Mathematics and the Chemical Engineering departments. Since 2006, he has been a Research Affiliate at the Laboratory for Information and Decision Systems, MIT, and since 2018 he has been the Faculty Member in the Program in Therapeutic Science at Harvard Medical School. His major current research interests lie in several areas of control and dynamical systems theory, systems molecular biology, cancer and immunology, and computational biology. Sontag has authored over five hundred research papers and monographs and book chapters in the above areas with about 59,000 citations and an h-index of 104 (51 since 2017). He is a Fellow of various professional societies: IEEE, AMS, SIAM, and IFAC, and is also a member of SMB and BMES. He was awarded the Reid Prize in Mathematics in 2001, the 2002 Hendrik W. Bode Lecture Prize and the 2011 Control Systems Field Award from the IEEE, the 2002 Board of Trustees Award for Excellence in Research from Rutgers, and the 2005 Teacher/Scholar Award from Rutgers.
 \end{IEEEbiography}
 
\end{document}